\documentclass[a4paper]{ar-1col}

\usepackage[utf8]{inputenc}
\usepackage{qcircuit}
\usepackage{amssymb}
\usepackage{amsmath}
\usepackage{colortbl}
\usepackage{siunitx}
\usepackage{upgreek}
\usepackage{multirow}
\usepackage{hyperref}
\hypersetup{
    colorlinks=true,
    linkcolor=red,
    citecolor=blue,
    filecolor=magenta,      
    urlcolor=blue,
}
\definecolor{fluxcolor}{RGB}{204, 217, 255}
\definecolor{uwavecolor}{RGB}{244, 220, 222}
\definecolor{FandUwavecolor}{RGB}{231,244,224}
\definecolor{cavitycolor}{RGB}{232, 200, 244}

\newcommand{\CZ}{\textsf{CZ}}

\begin{document}

% FRONT PAGE MATTER %
%%%%%%%%%%%%%%%%%%%%%
% Page header
\markboth{Kjaergaard et al.}{Superconducting Qubits: Current State of Play}

% Title
\title{Superconducting Qubits: Current State of Play}

%Authors, affiliations address.
\author{Morten Kjaergaard,$^1$ Mollie E. Schwartz,$^2$ Jochen Braum\"uller,$^1$ Philip Krantz,$^3$ Joel I-J Wang,$^1$, Simon Gustavsson,$^1$ and William D. Oliver,$^{1,2,4}$
\affil{$^1$Research Laboratory of Electronics, Massachusetts Institute of Technology, Cambridge, USA, MA 02139. MK email: mortenk@mit.edu}
\affil{$^2$MIT Lincoln Laboratory, 244 Wood Street, Lexington, USA, MA 02421}
\affil{$^3$Microtechnology and Nanoscience, Chalmers University of Technology, G\"oteborg, Sweden, SE-412 96}
\affil{$^4$Department of Physics, Massachusetts Institute of Technology, Cambridge, USA, MA 02139. WDO email: william.oliver@mit.edu}}

%Abstract
\begin{abstract}
Superconducting qubits are leading candidates in the race to build a quantum computer capable of realizing computations beyond the reach of modern supercomputers. The superconducting qubit modality has been used to demonstrate prototype algorithms in the `noisy intermediate scale quantum' (NISQ) technology era, in which non-error-corrected qubits are used to implement quantum simulations and quantum algorithms. With the recent demonstrations of multiple high fidelity two-qubit gates as well as operations on logical qubits in extensible superconducting qubit systems, this modality also holds promise for the longer-term goal of building larger-scale error-corrected quantum computers. In this brief review, we discuss several of the recent experimental advances in qubit hardware, gate implementations, readout capabilities, early NISQ algorithm implementations, and quantum error correction using superconducting qubits. While continued work on many aspects of this technology is certainly necessary, the pace of both conceptual and technical progress in the last years has been impressive, and here we hope to convey the excitement stemming from this progress.
\end{abstract}

%Keywords, etc.
\maketitle

% TEXT START %
%%%%%%%%%%%%%%

\section{INTRODUCTION} \label{sec:introduction}
% -------------------------------------------------------------------------------------
The ability to control individual quantum degrees of freedom and their interactions unlocks the capability to perform quantum coherent computation. This in turn imparts the possibility to perform certain computational tasks and quantum simulations which are outside the reach of modern supercomputers \cite{nielsen_quantum_2011,montanaro_quantum_2016}. Superconducting qubits -- collective excitations in superconducting circuits -- are currently one of the leading approaches for realizing quantum logic elements and quantum coherent interactions with sufficiently high controllability and low noise to be a viable candidate for implementing medium and large-scale quantum computation.

In 2014, the first controlled qubit-qubit interaction with fidelities greater than 0.99 in multi-qubit systems was demonstrated~\cite{barends_superconducting_2014} with the transmon qubit~\cite{Koch2007} variant of superconducting qubits, and since then, multiple controlled two-qubit interactions have been demonstrated with similarly high fidelities (e.g. Refs.~\cite{sheldon_procedure_2016} and~\cite{hong_demonstration_2019}). Even though the two-qubit gate fidelity in multi-qubit systems is a limited metric for evaluating the maturity of a quantum computing technology, it implies a high degree of control of all aspects of the quantum processor, and indicates the state of play: superconducting qubits are well positioned to be a platform for demonstrating interesting noisy intermediate-scale quantum computing (NISQ) \cite{preskill_quantum_2018} protocols outside the reach of classical computers and first realizations of operations on multiple logical error-corrected qubits \cite{gambetta_building_2017,devoret_superconducting_2013}.

\begin{figure}[!b]
\includegraphics[width=1\textwidth]{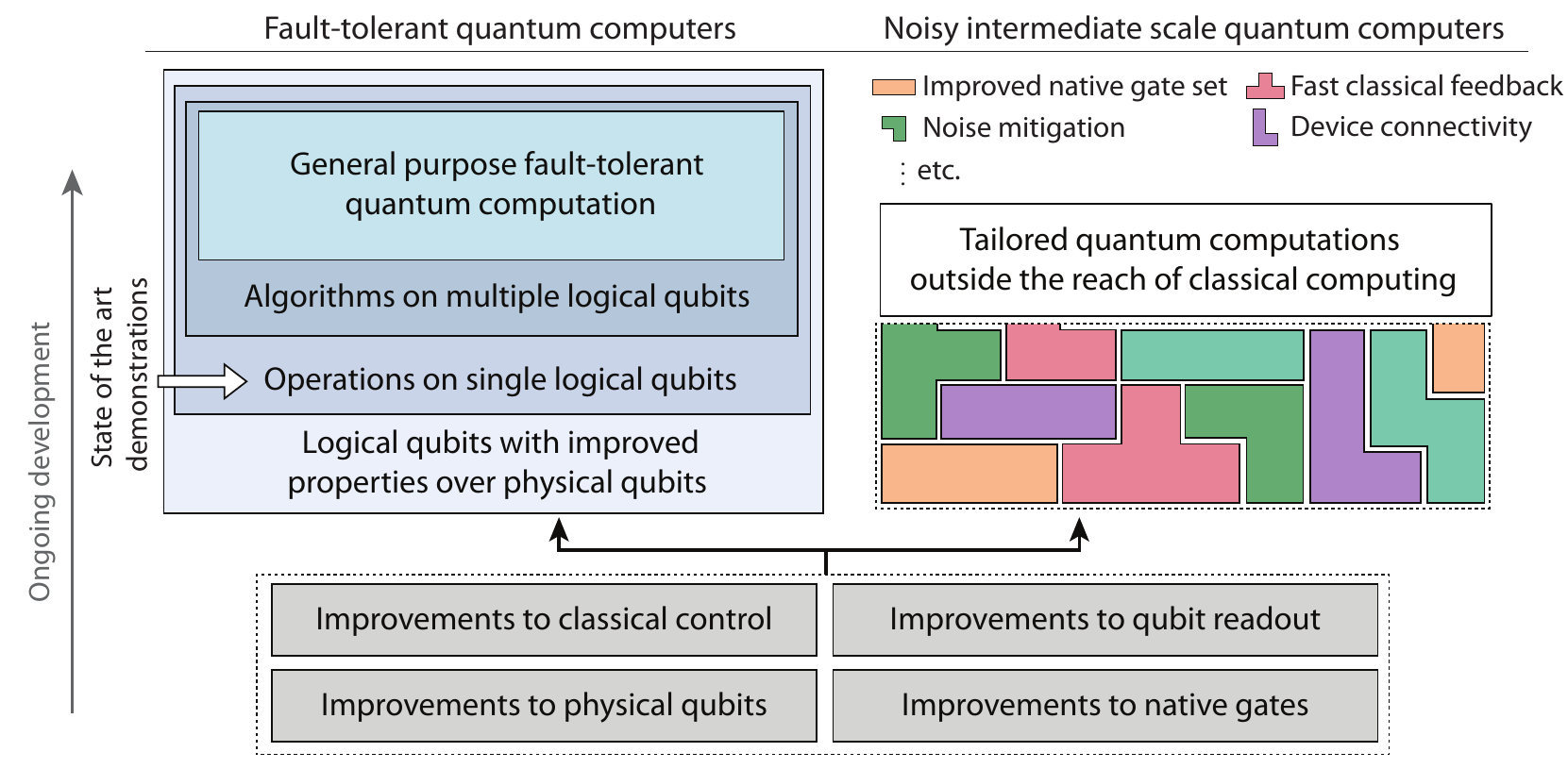}
\caption{Path towards fault-tolerant quantum error-corrected quantum computers (left) as well as noisy intermediate scale quantum computing (right) using superconducting qubits. The left track follows the path towards quantum computers capable of performing arbitrarily long programs to arbitrary precision, based on logical (i.e. encoded and error-corrected) qubits. The right track is the `NISQ' approach (see Ref.~\cite{preskill_quantum_2018}), where highly optimized quantum algorithms and quantum simulations, which typically take into account details of the quantum processor, can be executed without generalized quantum error correction procedures. The two tracks are pursued in parallel in many academic, government, and industrial laboratories.}
\label{fig:NISQ-QEC}
\end{figure}

In Fig.~\ref{fig:NISQ-QEC}, we show two major tracks being pursued in parallel in the community. The left track (see e.g. Refs.~\cite{devoret_superconducting_2013,jones_layered_2012}) shows the progression towards building a fault-tolerant quantum computer, capable of running an arbitrarily long computation, to arbitrary precision. Since 2012-2013, the field has climbed this ladder, including the recent demonstration of a logical qubit with a lifetime longer than any of the underlying constituent elements \cite{ofek_extending_2016}, operations on single logical qubits \cite{hu_demonstration_2018} as well as logical operations between two encoded (but not yet error-corrected) qubits \cite{rosenblum_cnot_2018}.

Although the architectures based purely on the transmon qubit (see Sec.~\ref{sec:SIS}) have not yet demonstrated logical states with longer lifetimes than their constituent physical states, multi-qubit systems with on the order of 10 -- 20 qubits have been demonstrated~\cite{kelly_state_2015,Otterbach2017,kandala_hardware-efficient_2017,neill_blueprint_2018,wei_verifying_2019}, and even larger systems with 50 -- 100 qubits are under current development~\cite{moore_ibm_2017,rigetti_rigetti_2018,hsu_ces_2018,kelly_preview_2018}. Such processors are eminently suitable for the NISQ era approach, where the aim is to perform quantum algorithms and quantum simulations that achieve quantum advantage in the absence of full quantum error correction. Such NISQ era demonstrations utilize highly tailored quantum programs that take into account detailed knowledge of a given quantum processor, some of which we have shown in the right track of Fig.~\ref{fig:NISQ-QEC}. These (and more) details will need to be combined judiciously to ultimately demonstrate non-trivial physics (i.e. quantum simulations) or calculations (i.e. quantum algorithms) in the NISQ approach. The full toolbox of the NISQ era is an area under active development, and the ultimate reach of this approach is not yet known. Prominent among the NISQ demonstrations is the recent result of demonstrating a clear quantum advantage (nicknamed `quantum supremacy'~\cite{harrow_quantum_2017}), where a computation performed on a quantum computer yields a result expected to be impossible to attain using large, classical supercomputers~\cite{neill_blueprint_2018,arute_supremacy_2019}.

In this review, we do not aspire to give a complete, chronological review of the entire field of superconducting qubits and their broad applicability for implementing circuit quantum electrodynamics (cQED) or as a platform for studying fundamental physics. Interested readers may consult any of the already existing excellent reviews (some of which can be found in e.g. Refs. \cite{Wendin2006, Clarke2008, girvin_circuit_2009,Oliver2013,devoret_circuit_2014,wendin_quantum_2017,gu_microwave_2017,krantz_quantum_2019,hauke_perspectives_2019}). Instead, we focus on highlights from each of the blocks in Fig.~\ref{fig:NISQ-QEC} that have brought the field to its current exciting state. In Section~\ref{sec:superconducting_hardware}, we review progress towards improving qubit coherence (Sec.~\ref{sec:SIS}), improved native gate fidelities (Sec.~\ref{sec:GatesControl}), improvements to readout (Sec.~\ref{sec:readout}) and developments in using resonators to act as quantum memories (Sec.~\ref{subsec:Resonator-based-encodings}). 

In Section~\ref{sec:NISQ_era_demos}, we review early NISQ-style demonstrations using superconducting qubits, including quantum supremacy (Sec.~\ref{sec:supremacy}), quantum simulation (Sec.~\ref{sec:qsimulation}), digital quantum algorithms (Sec.~\ref{sec:digitalquantumalgorithms}) and quantum annealing (Sec.~\ref{sec:quantumannealing}). In Section~\ref{sec:QEC}, we briefly introduce the framework of quantum error correction and review progress in experiments using parity readout, often used in the context of realizing subsections of the surface code (Sec.~\ref{subsec:surfacecode}), as well as experiments towards demonstrating fault-tolerance (Sec.~\ref{subsec:FT}) and operations on logical qubits encoded in resonator states (Sec.~\ref{subsec:catcodes}). Finally in Section~\ref{sec:outlook}, we provide an outlook on the developments from the preceeding sections, and discuss some of the near-term challenges related to moving to larger quantum processors based on the superconducting qubit modality.

% SECTION 2: QUBIT HARDWARE
% -------------------------------------------------------------------------------------
\section{THE HARDWARE OF SUPERCONDUCTING QUBITS}\label{sec:superconducting_hardware}
Superconducting circuits are manufactured using a multi-step additive and subtractive fabrication process involving lithographic patterning, metal deposition, etching, and controlled oxidation of thin two-dimensional films of a superconductor such as aluminum or niobium. Circuits are fabricated on silicon or sapphire substrates, leveraging techniques and materials compatible with silicon CMOS manufacturing. Devices are placed inside a copper or aluminum package that provides an engineered electromagnetic environment with requisite signal lines and thermally anchored to the $\approx\SI{10}{mK}$ stage of a dilution refrigerator. The toolbox of superconducting circuits comprises resonators and bias lines, in addition to the qubits themselves. The properties of these building blocks can be engineered by varying circuit parameters and interconnected with tailored couplings.
\begin{marginnote}[]
\entry{Josephson junction}{Superconducting qubits are based on the Josephson junction, which consists of two superconducting electrodes that are separated by a thin insulating barrier, allowing for the coherent tunneling of Cooper pairs, resulting in a lossless non-linear inductor.}
\end{marginnote}

\subsection{Devices based on superconducting tunnel junctions}\label{sec:SIS}
% -------------------------------------------------------------------------------------
\begin{figure}[t]
\includegraphics{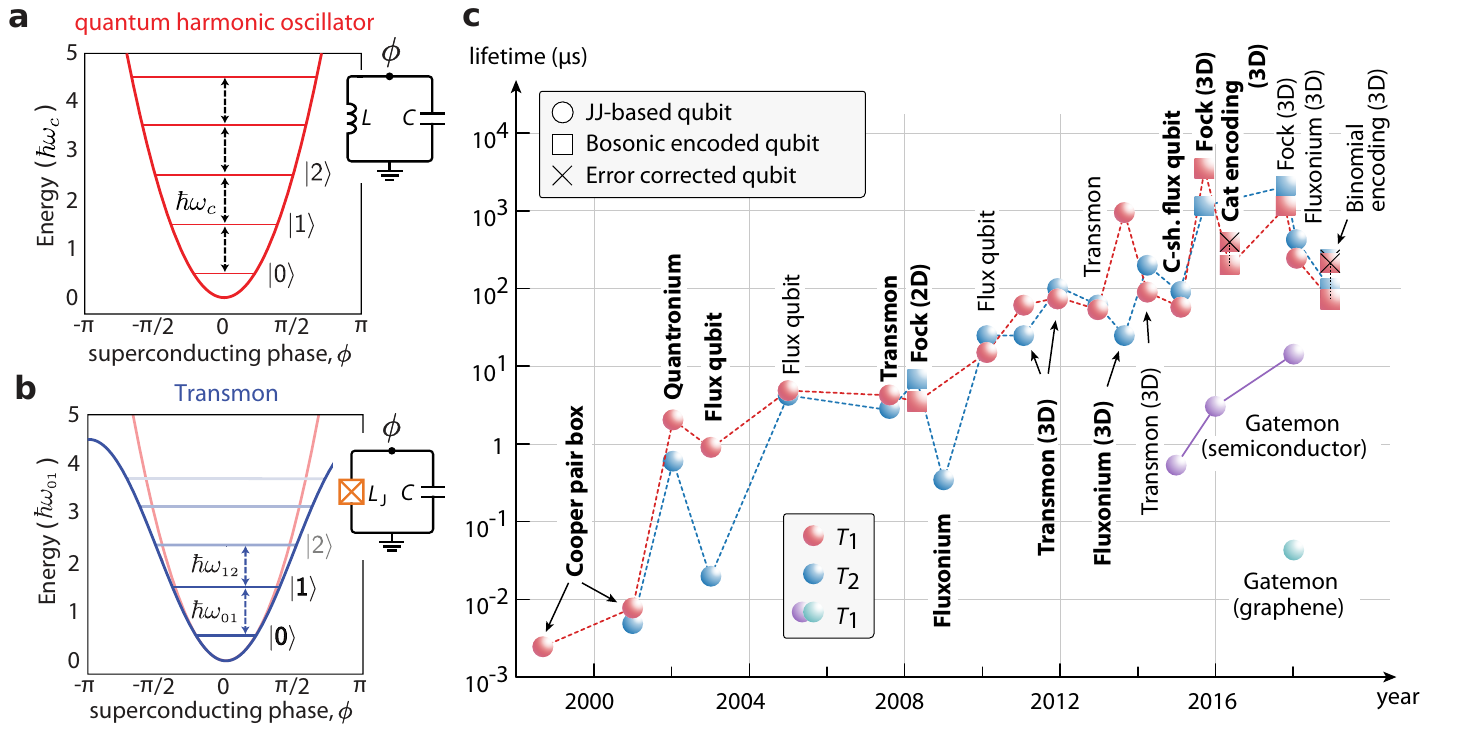}
\caption{
(a) The energy spectrum of a quantum harmonic oscillator (QHO).
(b) The energy spectrum of the transmon qubit, showing how the introduction of the non-linear Josephson junction produces non-equidistant energy levels.
(c) Evolution of lifetimes and coherence times in superconducting qubits. Bold font indicates the first demonstration of a given modality.
`JJ-based qubits' are qubits where the quantum information is encoded in the excitations of a superconducting circuit containing one or more Josephson junctions (see Sec.~\ref{sec:SIS}).
`Bosonic encoded qubits' are qubits where the quantum information is encoded in superpositions of multi-photon states in a QHO, and a Josephson junction circuit mediates qubit operation and readout (see Sec.~\ref{subsec:Resonator-based-encodings}).
`Error corrected qubits' represent qubit encodings in which a layer of active error-correction has been implemented to increase the encoded qubit lifetime.
The charge qubit and transmon modalities are described in Sec.~\ref{subsec:cpb_and_trans}, flux qubit and the capacitively shunted flux qubit (`C-sh. flux qubit') are described in Sec.~\ref{subsec:flux_and_cshuntflux}, and fluxonium and gatemon modalities are described in Sec.~\ref{sec:outlook}.
The codes underlying the `cat encoding' and `binomial encoding' are discussed in Sec.~\ref{subsec:catcodes}.
`(3D)' indicates a qubit embedded in a three-dimensional cavity.
For encoded qubits, the non-error-corrected $T_1$ and $T_2$ times used in this figure are for the encoded, but not error-corrected, version of the logical qubit (see Refs.~\cite{ofek_extending_2016} and~\cite{hu_demonstration_2018} for details).
The references for the JJ-based qubits are (in chronological order)~\cite{Nakamura1999,Nakamura2002,Vion2002,Chiorescu2003,Bertet2005,Houck2008,Manucharyan2009,Bylander2011,Paik2011,Rigetti2012,Chang2013,Pop2014,Jin2015,Yan2016,Nguyen2018}; the semiconductor-JJ-based transmons (gatemons) are Refs.~\cite{Larsen2015_1,Casparis2016,Luthi2018}; and the graphene-JJ-based transmon is Ref.~\cite{Wang2019}. The bosonic encoded qubits in chronological order are Refs.~\cite{hwang_2008,cwang_2016,ofek_extending_2016,rosenblum_fault-tolerant_2018,hu_demonstration_2018}.}
\label{fig:dispersion_devoret}
\end{figure}

The quantum harmonic oscillator (QHO) shown in Fig.~\ref{fig:dispersion_devoret}(a) is a resonant circuit comprising a capacitor and an inductor with resonance frequency $\omega_{\mathrm{c}}=1/\sqrt{LC}$. For sufficiently low temperature ($k_{\mathrm{B}}T \ll \hbar \omega_{\mathrm{c}})$ and dissipation (level broadening much less than $\hbar \omega_{\mathrm{c}}$), the resulting harmonic potential supports quantized energy levels spaced by $\hbar\omega_{\mathrm{c}}$. However, due to the equidistant level spacing, the QHO by itself cannot be operated as a qubit.

To remedy this situation, the circuit potential is made anharmonic by introducing a nonlinear inductor -- the Josephson junction. The imparted anharmonicity leads to a non-equidistant spacing of the energy levels, enabling one to uniquely address each transition, see Fig.~\ref{fig:dispersion_devoret}(b). Typically, the two lowest levels are used to define a qubit, with $|0\rangle$ corresponding to the ground state and $|1\rangle$ corresponding to the excited state. Large anharmonicity is generally favourable to suppress unwanted excitations to higher levels.

Beyond the simple circuit in Fig.~\ref{fig:dispersion_devoret}(b), one may add additional inductors, capacitors, and Josephson junctions to achieve certain design goals.
The shape of the potential energy landscape and the nature of the encoding of the qubit states (charge, flux, etc.) depend on the relative strengths of the energies associated with these various circuit elements, including the Josephson energy $E_{\mathrm{J}}$, the capacitive charging energy $E_{\mathrm{C}}$, and the inductive energy $E_{\mathrm{L}}$~\cite{devoret_superconducting_2013,Oliver2013,gu_microwave_2017}. Tuning the underlying circuit parameters enables one to engineer and trade-off various qubit properties, including transition frequency, anharmonicity, and sensitivity to various noise sources.

Contemporary superconducting circuits evolved from two fundamental types of qubits: one based on electric charge and one based on magnetic flux. These initial modalities -- charge and flux qubits, respectively -- have been improved and generalized over the past 20 years to realize the multiple types of qubits in use today~\cite{Wendin2006,devoret_circuit_2014,krantz_quantum_2019}.

\subsubsection{Charge qubits and derivatives}\label{subsec:cpb_and_trans} The first temporal coherence in a superconducting circuit was observed in a charge qubit (also called a Cooper-pair box)~\cite{Nakamura1999}. A charge qubit consists of a small superconducting island connected to a large superconducting reservoir via a Josephson junction. A capacitively coupled gate voltage controls the charge offset $n_{\mathrm{g}}$ on the island, and it is used to tune the qubit frequency. The circuit Hamiltonian is given by
\begin{equation}
\hat H=4E_{\mathrm{C}}\left(\hat N- n_{\mathrm{g}}\right)^2 -E_{\mathrm{J}}\cos\hat\phi,
\label{eq:charge_qubit}
\end{equation}
where $\hat N$ denotes the number of excess Cooper pairs on the island, $\hat\phi$ is the $2\pi$-periodic operator of the phase difference across the Josephson junction, and the operators satisfy the commutation relation $[\hat\phi, \hat N]=i$.
Charge qubits are designed in the regime $E_\textrm{C} > E_\textrm{J}$, such that the island charge is a good quantum number. The bare qubit states are $|N\rangle$ and $|N+1\rangle$, corresponding to the absence and presence of an additional Cooper pair on the island. The Josephson junction acts as a valve for Cooper pairs and couples these states, opening an avoided crossing of size $E_\textrm{J}$ at integer multiples of offset charge $n_\textrm{g} = 1/2$. Although charge qubits have large anharmonicity $\alpha \equiv \omega_{12}/2\pi - \omega_{01}/2\pi > \SI{10}{GHz}$, their lifetime and dephasing is strongly limited by environmental charge noise~\cite{Oliver2013}. In addition, the small size of the island and Josephson junction leads to a strong susceptibility to stray capacitance, local defects, and fabrication variation, leading to large device-to-device variability.

To mitigate these issues, a large shunt capacitor was added to the charge qubit -- a device nicknamed ``the transmon''~\cite{Koch2007}. The transmon is also described by Eq.~\eqref{eq:charge_qubit}, but it is designed in the regime $E_\mathrm{J}/E_\mathrm{C}\gtrsim 50$, resulting in an exponential reduction of its charge noise sensitivity and making it a ``charge-insensitive charge qubit'' (charge is no longer a good quantum number). Adding a high-quality shunt capacitor has led to improved reproducibility and coherence times in the range of \SIrange{50}{100}{\micro s}, see Fig.~\ref{fig:dispersion_devoret}(c).
%
% margin notes
%
\begin{marginnote}[]
\entry{Cooper pair box}{A charge qubit whose states correspond to the presence or absence of an additional Cooper pair on a superconducting island. Although the first superconducting qubit to exhibit temporal coherence, it suffered from poor reproducibility and coherence times.}
\entry{Transmon qubit}{A capacitively shunted variant of the Cooper pair box that is largely insensitive to charge, resulting in improved reproducibility and coherence times. It is one of the leading modalities used today for gate-model quantum computing.}
\entry{SQUID}{A superconducting quantum interference device consists of a superconducting closed loop that is interrupted by one (rf-SQUID) or two (dc-SQUID) Josephson junctions. It is employed in superconducting circuits to enable frequency tunability via an applied magnetic flux.}
\end{marginnote}

The improved performance of the transmon comes at the expense of a reduced anharmonicity to values of about $\SI{-200}{MHz}$~\cite{Koch2007}, being only a few percent of the qubit level spacing $\omega_\text{q}/2\pi \equiv \omega_{01}/2\pi \sim\SI{5}{GHz}$. For single-junction transmons (see Fig.~\ref{fig:dispersion_devoret}(b), this frequency is set by the size of the shunt capacitor and the critical current $I_\text{c}$ of the Josephson junction, determined by design and fabrication parameters such as materials choice, junction area, and insulator thickness. Replacing the single Josephson junction by a superconducting loop with two junctions in parallel -- a dc-SQUID -- enables one to tune the effective critical current of the Josephson junction (and hence the qubit frequency) via a magnetic field applied to the dc-SQUID loop. The trade-off for this additional control knob is that the qubit becomes susceptible to magnetic flux noise.

Transmon qubits can be coupled capacitively -- either directly or as mediated by a resonator ``bus'' -- which, in the natural eigenbasis of the transmon qubits, lead to a two-qubit interaction term of the form $\hat H^\text{int}_\text{cap} = J_\text{cap}\left(\hat \sigma_x ^i \hat\sigma_x ^j + \hat\sigma_y ^i\hat\sigma_y ^j\right)$. The physical coupling strength $J_{\mathrm{cap}}$ is related to the coupling capacitance and in the case of a resonator bus, the frequency detuning between the qubits and the resonator. By introducing an additional ``coupler qubit'' or ``coupler SQUID'', one can furthermore tune the effective coupling strength~\cite{chen_qubit_2014}.

\subsubsection{Flux qubits and derivatives} \label{subsec:flux_and_cshuntflux}
The superconducting qubit modality based on magnetic flux underwent a similar evolution. With flux qubits, the bare qubit states are defined by circulating currents in a superconducting loop interrupted by a small Josephson junction in series with either a linear inductor or several larger-area Josephson junctions. The small junction acts as a valve for magnetic fluxons, allowing one to enter or leave the superconducting loop. The presence or absence of this fluxon is accompanied by a clockwise or counterclockwise circulating current, which serves to satisfy the flux quantization condition in concert with the Josephson junctions, the linear inductances, and an externally applied magnetic field threading the loop. When biased at half a flux quantum, the small junction couples and hybridizes these states with a strength related to $E_{\mathrm{J}}$, $E_{\mathrm{C}}$, and $E_{\mathrm{L}}$.

Within a two-level approximation, the flux qubit potential is approximated by two wells of energy difference $\epsilon$ and coupling energy $\Delta$, yielding an effective two-level Hamiltonian $\hat H _{\mathrm{TL}} = \epsilon(\Phi_z)\hat\sigma_z +\Delta\hat\sigma_x$~\cite{Wendin2006,Clarke2008}, with a qubit frequency $\hbar\omega_{\mathrm{q}}=\sqrt{\epsilon^2+\Delta^2}$, and $\Phi_z$ is the flux applied to the flux qubit loop (typically denoted the $z$-loop). By replacing the small Josephson junction with a secondary dc-SQUID loop, the coupling $\Delta$ becomes flux-tunable, leading to the modified two-level Hamiltonian $\hat H' _{\mathrm{TL}} = \epsilon(\Phi_z,\Phi_x)\hat\sigma_z +\Delta(\Phi_x)\hat\sigma_x$, where $\Phi_x$ is the flux applied to the $x$-loop. This makes the flux qubit a spin-1/2 system with tunable $z$ and $x$ fields, a building block for quantum annealing applications based on the transverse Ising Hamiltonian \cite{hauke_perspectives_2019}.

%
% margin notes
%
\begin{marginnote}[]
\entry{Flux qubit}{A qubit modality based on magnetic flux whose states correspond to clockwise and counter-clockwise currents flowing around a loop interrupted by Josephson junctions. %Used widely for quantum annealing.
Although the persistent-current flux qubit exhibited a high degree of temporal coherence, it suffered from poor reproducibility.}
\entry{Capacitively shunted flux qubit}{A capacitively shunted variant of the persistent-current flux qubit used for both gate-model and quantum annealing circuits. It features improved reproducibility and coherence times while retaining $>\SI{500}{MHz}$ anharmonicity.}
\end{marginnote}

In the context of gate-model quantum computing, the persistent-current flux qubit~\cite{Mooij1999,Orlando1999} was the most successful of the early flux qubits, featuring a small junction (the valve) in series with 2 or 3 larger-area Josephson junctions (the series inductance). As with the transmon that later followed, this qubit operates in the regime $E_{\mathrm{J}}\gg E_{\mathrm{C}}$ and is largely charge-insensitive. In addition, it featured a large anharmonicity with moderately-high coherence times~\cite{Bertet2005}, including the first superconducting qubit demonstrating coherence exceeding $\SI{10}{\micro s}$ (Fig.~\ref{fig:dispersion_devoret}(c)) and reaching as high as
$\SI{23}{\micro s}$~\cite{Bylander2011}. However, like the charge qubit, its major limitation was a lack of device-to-device reproducibility.

To improve the flux qubit, a large shunt capacitance was again added~\cite{Yan2016,Nori2007,Steffen2010}. The resulting ``capacitively shunted flux qubit'' featured improved reproducibility at the expense of qubit anharmonicity, in this case to around $\SI{500}{MHz}$. It also reduced the circulating current, resulting in reduced sensitivity to flux noise and leading to coherence times in the range of \SIrange{50}{100}{\micro s} (see Fig.~\ref{fig:dispersion_devoret}(c)).

Flux qubits generally are coupled inductively to each other, resulting in an interaction term of the form $\hat{H}^\text{int}_\text{ind} =J_\text{ind}\hat\sigma_z ^i\hat\sigma_z ^j$. The coupling strength $J_\text{ind}$ can be tuned by the magnetic flux applied to an additional inductive coupling element~\cite{Harris2009,Weber2017}, with the potential to implement noise-resilient two-qubit gates~\cite{kerman_high-fidelity_2008}.

\subsubsection{Qubit modalities -- the current state of play}
The transmon is currently the most widely used qubit for gate-based quantum computation, and it has been used to demonstrate multiple high-fidelity logical operations, quantum simulations and digital algorithms (see Sec.~\ref{sec:GatesControl}, Sec.~\ref{sec:qsimulation} and Sec.~\ref{sec:digitalquantumalgorithms}).
In turn, due to the structure of their Hamiltonians, the persistent-current and rf-SQUID flux qubits are currently the predominant platforms being used for quantum annealing (see Sec.~\ref{sec:quantumannealing}), including the commercial D-Wave system~\cite{Johnson2011}.
With the advent of capacitively shunted flux qubits, this modality now also supports high reproducibility, long coherence times, and moderate anharmonicity levels. Combined with the tunability of its Hamiltonian, this qubit offers a potential alternative platform for Hamiltonian emulation, gate-based quantum computing and quantum annealing.

Today, a ``generalized superconducting qubit'' framework is emerging, featuring a capacitively shunted small junction in series with $N$ larger-area Josephson junctions (or an inductive shunt). The transmon is an early example of this evolution, as is the capacitively shunted flux qubit.
Another example is the fluxonium qubit~\cite{Manucharyan2009} (see also Sec.~\ref{sec:outlook}), which has been demonstrated with coherence times exceeding \SI{100}{\micro s}~\cite{Pop2014,Nguyen2018} at the expense of increased complexity in the number of Josephson junctions.

% -------------------------------------------------------------------------------------
\subsection{Gate operations in superconducting qubits}\label{sec:GatesControl}
The predominant technique for implementing single-qubit operations is via microwave irradiation of the superconducting circuit. Electromagnetic coupling to the qubit with microwaves at the qubit transition frequency drives Rabi oscillations in the qubit state. Control of the phase and amplitude of the drive is then used to implement rotations about an arbitrary axis in the $x,y$ plane. Within the rotating wave approximation, a microwave drive resonant with the qubit frequency gives rise to the Hamiltonian $\hat H_\text{drive} = \Omega \left( I(t)\hat\sigma_x + Q(t)\hat\sigma_y\right)$, where $I(t)$ ($Q(t)$) is the envelope function of the in-phase (quadrature) component of the microwave signal and $\Omega$ is the Rabi frequency as experienced by the qubit. However, due to the typically low anharmonicity of the transmon qubit, higher-order levels are easily populated, leading to leakage and dephasing effects.\begin{marginnote}[]
\entry{I,Q}{$I(t)$ and $Q(t)$ are the in-phase and quadrature components of the amplitude of the microwave drive. The labels are borrowed from classical RF processing.}\end{marginnote}
To counteract this, the Derivative Removal by Adiabatic Gate (DRAG) technique is typically used to enable fast gates without leakage into higher-level states \cite{motzoi_simple_2009}, and single-qubit gates are now routinely implemented with fidelities $\gtrsim0.99$~(e.g. Ref.~\cite{barends_superconducting_2014,gustavsson_improving_2013,sheldon_characterizing_2016,rol_restless_2017,reagor_demonstration_2018}), typically measured using interleaved Clifford randomized benchmarking \cite{magesan_scalable_2011}). $z$-axis rotations are typically performed in a virtual manner, where the phase of the qubit drives are shifted, effectively producing a $z$-rotation \cite{mckay_efficient_2017}.

While the implementation of single-qubit gates is now mostly uniform across the community, many different two-qubit gates have been demonstrated, and several of those have reached fidelities $>0.99$. \begin{marginnote}[]
\entry{Interleaved Clifford randomized benchmarking}{A technique for assessing the average fidelity of a quantum gate, by interleaving the gate of interest in sequences of Clifford gates, and randomizing over many such sequences.}
\end{marginnote}The two-qubit gates can be roughly split into three categories. One class uses tunable transmon qubits whose frequencies can be modulated by applying magnetic flux through a dc-SQUID loop that tunes the effective critical current of the Josephson junction. Several high-fidelity two-qubit gates can be implemented by tuning certain transitions close to resonance \cite{barends_superconducting_2014,chen_qubit_2014,dicarlo_demonstration_2009,kjaergaard_2019,dewes_characterization_2012} (see details in Tbl.~\ref{tab:2qubitgates}). The second class uses fixed-frequency qubits which are manipulated by microwave irradiation, typically driving one qubit at the frequency of a second qubit, to enact high-fidelity entangling gates \cite{chow_simple_2011,poletto_entanglement_2012,chow_microwave-activated_2013,paik_experimental_2016}. The third class relies on parametrically driving a coupling element (or the qubits themselves) to induce a tunable coupling between the qubits. Such operations are referred to as "parametrically driven", and two high-fidelity two-qubit gates have recently been demonstrated using such parametrically driven interactions \cite{hong_demonstration_2019,mckay_universal_2016,caldwell_parametrically_2018}. Common to all these gates is that they generate entanglement in the system via conditional rotations or transitions, such that the state and/or the phase of one qubit becomes dependent on that of the other. The class that uses tunable qubits has increased sensitivity to flux noise, but gates can be implemented more quickly. Conversely, the fixed frequency devices typically have longer lifetimes, but also require longer gate operation times. Table~\ref{tab:2qubitgates} shows the current state-of-the-art fidelities of the two-qubit gates demonstrated to date. The continued development of novel gate designs, and fidelity improvement in current designs is a highly active area of research.

\begin{table}[!t]
\caption{State of the art high-fidelity two-qubit gates in superconducting qubits}
\label{tab:2qubitgates}
\vspace*{1mm}
\hspace*{-2.89cm}\begin{tabular}{lcllcl}
\hline
Acronym$^{\rm a}$							&  	Layout$^{\rm b}$			& First demonstration [Year] 	& Highest fidelity [Year]	&	Gate time \\
\hline
\cellcolor{fluxcolor} 										& 	\cellcolor{fluxcolor} 								& \cellcolor{fluxcolor} 		&\cellcolor{fluxcolor}99.4\%$^{\rm \dagger}$ Barends et al. \cite{barends_superconducting_2014} [2014]&\cellcolor{fluxcolor}  $\SI{40}{ns}$ \\
\multirow{-2}{*}{\cellcolor{fluxcolor}\textsf{CZ} (ad.) } 	&  \multirow{-2}{*}{\cellcolor{fluxcolor}T--T}	& \multirow{-2}{*}{\cellcolor{fluxcolor}DiCarlo et al. \cite{dicarlo_demonstration_2009}  [2009]} &\cellcolor{fluxcolor}99.7\%$^{\rm \dagger}$ Kjaergaard et al. \cite{kjaergaard_2019} [2020]& \cellcolor{fluxcolor} $\SI{60}{ns}$ \\
\rowcolor{fluxcolor}$\sqrt{\textsf{iSWAP}}$	& T--T			& Neeley et al. \cite{neeley_generation_2010}$^{\rm \circ}$  [2010]& 90\%$^{\rm \star}$ \:\:\:Dewes et al. \cite{dewes_characterization_2012} [2014] 		 & $\SI{31}{ns}$ \\
\rowcolor{uwavecolor}\textsf{CR} 		 	& F--F			& Chow et al. \cite{chow_simple_2011}  [2011] 						& 99.1\%$^{\rm \dagger}$ Sheldon et al. \cite{sheldon_procedure_2016} [2016]		 & $\SI{160}{ns}$ \\
\rowcolor{uwavecolor}$\sqrt{\textsf{bSWAP}}$& F--F	 		& Poletto et al. \cite{poletto_entanglement_2012}  [2012]			& 86\%$^{\rm \star}$ \:\: ibid. 		& $\SI{800}{ns}$ \\
\rowcolor{uwavecolor}\textsf{MAP}		 	& F--F 			& Chow et al. \cite{chow_microwave-activated_2013} [2013] 			& 87.2\%$^{\rm \star}$ ibid.  & $\SI{510}{ns}$ \\
\rowcolor{fluxcolor}\textsf{CZ} (ad.) 		 & T--(T)--T & Chen et al. \cite{chen_qubit_2014}  [2014] 					& 99.0\%$^{\rm \dagger}$ ibid. 		 & $\SI{30}{ns}$ \\
\rowcolor{uwavecolor}\textsf{RIP}	 		 & 3D F 	& Paik et al. \cite{paik_experimental_2016}  [2016]				& 98.5\%$^{\rm \dagger}$ ibid. 	 	& $\SI{413}{ns}$ \\
\rowcolor{FandUwavecolor}$\sqrt{\textsf{iSWAP}}$ & F--(T)--F & McKay et al. \cite{mckay_universal_2016} [2016]			& 98.2\%$^{\rm \dagger}$ ibid. 		& $\SI{183}{ns}$ \\
\rowcolor{FandUwavecolor}\textsf{CZ} (ad.)	 	& T--F		& Caldwell et al. \cite{caldwell_parametrically_2018}  [2018]	& 99.2\%$^{\rm \dagger}$ Hong et al. \cite{hong_demonstration_2019} [2019]		& $\SI{176}{ns}$ \\
\rowcolor{cavitycolor}\textsf{CNOT}$_L$  		& BEQ-BEQ 	& Rosenblum et al. \cite{rosenblum_cnot_2018}  [2018] 			& $\sim$99\%$^{\rm \square}$ ibid. & $\SI{190}{ns}$ \\
\rowcolor{cavitycolor}\textsf{CNOT}$_{T-L}$  & BEQ-BEQ & Chou et al. \cite{Chou2018}  [2018]						 &	79\%$^{\rm \star}$\:\:\:\: ibid.& $\SI{4.6}{\micro s}$ \\
\hline
\end{tabular}

\vspace*{1mm}
\begin{tabnote}
Gates ordered by year of first demonstration. Gate time is for the highest fidelity gate.\\
$^{\rm a}$Full names: \textsf{CZ} (ad.): Adiabatic controlled phase, $\sqrt{\textsf{iSWAP}}$: square-root of the \textsf{iSWAP}, \textsf{CR}: Cross-resonance, $\sqrt{\textsf{bSWAP}}$: Square-root of the Bell-Rabi SWAP, \textsf{MAP}: Microwave activated phase, \textsf{RIP}: Resonator induced phase gate, \textsf{CNOT}$_L$: Logical \textsf{CNOT}, \textsf{CNOT}$_{T-L}$: Teleported logical \textsf{CNOT}.
\\$^{\rm b}$F is short `fixed frequency', T is short for `tunable'. For all non-bosonic encoded qubit gates, the qubits were of the transmon variety (except for the first demonstration of $\sqrt{\textsf{iSWAP}}$, using phase qubits, and first demonstration of $\textsf{CR}$ which used capacitively shunted flux qubits). Terms in parenthesis is a coupling element. `3D F' is short for a fixed frequency transmon qubit in a three-dimensional cavity. `BEQ' is short for bosonic encoded qubit (see Sec.~\ref{subsec:Resonator-based-encodings}).
\\$^{\rm \circ}$Implemented with phase qubits.
\\$^{\rm \dagger}$Determined by interleaved randomized Clifford benchmarking \cite{magesan_scalable_2011}.
\\$^{\rm \square}$Determined by repeated application of the gate to various input states and observing state fidelity decay as function of applied gates. See~\cite{rosenblum_cnot_2018} for details.
\\$^{\rm \star}$Determined by quantum process tomography.
\\
\raisebox{-0.2em}{\textcolor{fluxcolor}{\rule{1em}{1em}}} Gates implemented on flux-tunable qubits.\\ 
\raisebox{-0.2em}{\textcolor{uwavecolor}{\rule{1em}{1em}}} All-microwave gates. \\
\raisebox{-0.2em}{\textcolor{FandUwavecolor}{\rule{1em}{1em}}} Combination of tunable and fixed frequency components. \\
\raisebox{-0.2em}{\textcolor{cavitycolor}{\rule{1em}{1em}}} Gates on bosonic encoded qubits.
\end{tabnote}
\end{table}
\subsection{Amplification and high-fidelity readout}\label{sec:readout}
% -------------------------------------------------------------------------------------
An essential part of any superconducting quantum chip is fast and reliable readout of its qubit states. For superconducting qubits, readout is typically done using dispersive readout, in which each qubit is entangled with photons in a linear readout resonator with frequency $\omega_\text{r}$ \cite{Blais2004,Wallraff2004}.

In the dispersive regime, when the qubit-resonator detuning $\Delta = \omega_\text{q} - \omega_\text{r}$ is much larger than their coupling rate $g$, no direct exchange of energy takes place between the two systems. Instead, the qubit and resonator shift each others' frequencies -- proportional to their photon occupations, coupling strength $g$, and detuning $\Delta$ -- enabling the readout of the qubit state by probing the microwave response of the resonator. Depending on the state of the qubit, the readout resonance shifts by the dispersive shift $\chi$. For a two-level system, $\chi$ is given by $g^2/\Delta$, and for the transmon qubit $\chi$ is modified to $\frac{g^2}{\Delta}\frac{\alpha}{\alpha+\Delta}$ (valid in the transmon regime, where $\alpha = -E_\text{c}$) \cite{Koch2007}. For efficient readout, $\kappa$ is designed to be similar to $\chi$, typically in the range of few $\SI{}{MHz}$. \begin{marginnote}[]
\entry{Dispersive coupling Hamiltonian}{Describes coupling of a resonator ($\hat a$) and a qubit ($\hat\sigma_z$). A canonical transformation of the Jaynes - Cummings Hamiltonian to second order in $g/\Delta$ yields the dispersive coupling term,\\  $\chi\hat{a}^\dagger\hat{a}{\hat \sigma}_z$, where $\chi$ is the dispersive coupling coefficient.}
\end{marginnote}While an increased $\kappa$ decreases the resonator ring-up time and thereby provides fast qubit-state readout, the coherence time of the qubit is increasingly limited by spontaneous energy decay into the readout cavity mode, referred to as the Purcell effect~\cite{Houck2008}. To mitigate this, the community is using so-called ``Purcell filters'', which essentially act as bandpass filters, that support strong interactions between the resonator and an output line, while protecting the qubit from energy decay~\cite{Reed2010}.

Dispersive readout requires relatively low photon numbers, which must be amplified with high quantum efficiency in order to enable fast, high-fidelity single-shot readout for real-time quantum feedback \cite{Clerk2010}. This requirement has motivated the development of quantum-limited parametric amplifiers \cite{Yurke1988,Yamamoto2008,Simoen2015} and detectors \cite{Mallet2009,Vijay2009,Lin2014,Krantz2016}.
\begin{marginnote}[]
\entry{Purcell effect}{Qubit decay into a nearby oscillator mode. In the absence of a Purcell filter, $\Gamma_\text{p} \approx (g/\Delta)^2\kappa$}
\end{marginnote}
Current state-of-the-art processors utilize frequency-multiplexed readout circuits, reducing the hardware overhead by coupling several readout resonators to the same amplifier chain \cite{barends_superconducting_2014}. The number of readout resonators that can be multiplexed is often limited by the bandwidth and saturation power of the parametric amplifier -- a limitation that has motivated the development of stepped-impedance parametric amplifiers with increased bandwidth~\cite{Mutus2014,Roy2015}, as well as Josephson traveling wave parametric amplifiers (JTWPAs), achieving both large bandwidth and high saturation power~\cite{Macklin2015}.

\subsection{Bosonic encoded qubits} \label{subsec:Resonator-based-encodings}
% -------------------------------------------------------------------------------------
Bosonic encoded qubits, or qubits encoded in the infinite-dimensional Hilbert space of a quantum harmonic oscillator (QHO), are in some ways the inverse of the technology described to this point. The encoding is defined by superpositions of multi-photon states in the QHO, whose modes obey bosonic statistics. The QHO is typically realized by either an engineered electromagnetic mode in a 3D microwave cavity or a lithographically defined transmission line resonator on a 2D chip. In these qubit implementations, the QHO encodes a qubit and is coupled to a transmon that plays a supplementary role in control and readout. The lack of individually-addressable energy level transitions in a QHO makes qubit manipulation more difficult than for transmons, but universal control is achieved using microwave irradiation and manipulations of the coupled transmon~\cite{Krastanov2015}. These encodings are attractive because they take advantage of the long lifetimes of microwave cavities \cite{Reagor2013,Reagor2016} and may enable hardware-efficient quantum error correction~\cite{Gottesman2001} (QEC). Significant recent effort has led to demonstrations of resonator state manipulation \cite{Heeres2015,Heeres2017} and readout \cite{Sun2014,rosenblum_fault-tolerant_2018} schemes, which have been used to demonstrate fault-tolerant measurements, error detection and correction, and active and passive QEC (see Sec.~\ref{subsec:catcodes}).

\begin{marginnote}[]
\entry{Coherent state $|{\alpha}\rangle$}{A minimum-uncertainty state of a QHO, comprised of a Poisson distribution of Fock states. Its is parametrized by an average photon number $\overline{n} = |\alpha|^2$ and a complex phase $\phi = \arg(\alpha)$. Coherent states are eigenstates of the ladder operator $\hat{a}$.}
\entry{Fock state $|{n}\rangle$}{A state of the QHO characterized by a single, well-defined photon occupation number $n$.}
\end{marginnote}
The bosonic encoding is implemented in superconducting hardware by coupling a long-lived microwave resonator to a transmon qubit which is additionally coupled to an auxiliary resonator that is used to read out the state of the transmon qubit. For a bosonic mode coupled to a transmon qubit with $\chi/\kappa \gg 1$, the dispersive coupling imparts a well-resolved photon-number dependent shift in the transmon frequency: $\omega_{q, n} = \omega_q - n\chi$. This is known as the "photon-number resolved regime" \cite{Gambetta2006,Schuster2007}. Photon state manipulations take two general forms: displacements $\hat{\mathcal{D}}(\alpha)$ that coherently add or remove energy, and selective number-dependent arbitrary phase (SNAP) operations $\hat{S}(\vec{\theta})$ that add an arbitrary phase to individual Fock states. Krastanov et al. showed that the combination of displacements and SNAP gates provides universal control over the resonator state \cite{Krastanov2015}.
\begin{marginnote}[]
\entry{Displacement operator}{$\hat{\mathcal{D}}(\alpha) = \textrm{exp}(\alpha \hat a^\dagger-\alpha^* \hat a)$}
\entry{SNAP operator}{$\hat{S}(\vec{\theta}) = \sum_{n=0}^\infty e^{i\theta_n} |{n}\rangle\langle{n}|$}
\end{marginnote}

Displacement operations are native to the QHO and are accomplished by applying a microwave drive to a weakly-coupled port at the resonator frequency. SNAP operations, because they address single energy levels within the QHO, require nonlinearity, and are realized using the photon-number resolved regime to entangle the transmon with the resonator and manipulate individual Fock states. Applying a slow pulse to the transmon qubit at frequency $\omega_{q, n}$ with $\tau_{\textrm{pulse}} \gg 1/\chi$ ensures that the bandwidth of the pulse is smaller than the spacing between the various $\omega_{q, n}$. In this case, the transmon qubit will be selectively pulsed \textit{if and only if} the resonator is in Fock state $|{n}\rangle$. The selective drive is then designed to impart a geometric phase to the resonator $|{n}\rangle$ state. By applying superposed drives at multiple $\omega_{q, n}$, an arbitrary geometric phase is imparted to each Fock state, thus implementing an arbitrary SNAP gate in a single step \cite{Heeres2015}.

The dispersive interaction between the transmon qubit and the resonator also enables readout of the parity of the resonator state \cite{Sun2014}. Here, parity refers to the symmetry of the coherent superposition(s) in the resonator: for example, the states $|\alpha\rangle \pm |-\alpha\rangle$ have parity $P = \pm 1$. Parity readout is particularly useful because the most common bosonic QEC codes use parity flips as an error syndrome (see Sec. \ref{subsec:catcodes}). The parity readout technique can further be used to reconstruct the full Wigner function of the resonator state \cite{Lutterbach1997}. Finally, fault-tolerant approaches to resonator parity measurement have been proposed \cite{Cohen2017} and demonstrated \cite{rosenblum_fault-tolerant_2018}.

% SECTION 3: NISQ DEMONSTRATIONS
% -------------------------------------------------------------------------------------
\section{EARLY NISQ ERA DEMONSTRATIONS USING SUPERCONDUCTING QUBITS} \label{sec:NISQ_era_demos}
In this section, we discuss NISQ computing implementations, which operate on noisy quantum hardware in the absence of quantum error correction. Recent demonstrations in this so-called NISQ era seek to perform useful quantum computations while tolerating some system noise in order to stretch limited (intermediate-scale) quantum resources to their maximum effect. NISQ demonstrations are mostly at the proof-of-principle stage and, with one notable exception—sampling solutions of a random circuit—generally have not outperformed a large classical computer in wall-clock time or accuracy. However, a computational advantage seems in reach for many of the experiments discussed below, by scaling up the number and quality of qubits on the chip and, consequently, the problem size. Although the task of controlling enough qubits to perform nontrivial demonstrations remains a major technological challenge, it is believed that on the order of 50–70 qubits with sufficiently high fidelities can achieve this goal~\cite{boixo_characterizing_2018}.

We organize this section into four branches of early NISQ-era implementations with soft borders: NISQ-Era Platforms and a Demonstration of Quantum Supremacy (Section~\ref{sec:supremacy}) discusses the recent report of quantum supremacy on a 53-qubit processor and the important role online cloud-based quantum computers will play in advancing NISQ-era algorithms. Quantum Simulations with Superconducting Circuits (Section~\ref{sec:qsimulation}) use a physical quantum system in order to study another quantum system of interest. Whereas errors in the physical qubits decrease the simulation fidelity, meaningful results can be extracted, e.g., if the timescale of interest is small compared with the decay times of the participating qubits. In contrast, digital quantum simulations and, more generally, universal digital algorithms are gate–based approaches that harness the power of a quantum processor to solve a problem that need not be quantum in nature (Section~\ref{sec:digitalquantumalgorithms}). Typically, the latter algorithms are tailored to specific (potentially noisy) hardware in order to maximize the overall fidelity of the computation. The fourth flavor is quantum annealing (Section~\ref{sec:quantumannealing}), representing a potential complementary approach to quantum computation.

\subsection{NISQ-Era Platforms and a Demonstration of Quantum Supremacy}\label{sec:supremacy}
% -------------------------------------------------------------------------------------
Developing commercializable NISQ-era algorithms will rely on access to quantum computers of increasing complexity and quality. Cloud-based access to superconducting quantum systems enables algorithm designers and others with expertise outside a traditional physics or quantum hardware background to try ideas. The approach was pioneered at IBM with a 5-transmon qubit device in 2016~\cite{IBM_press_release}. As of the writing of this review, such access is available from IBM, Rigetti Computing, and D-Wave (see, respectively, \href{https://www.ibm.com/quantum-computing/}{https://www.ibm.com/quantum-computing/}, \href{https://www.rigetti.com/qcs}{https://www.rigetti.com/qcs}, and \href{https://cloud.dwavesys.com/}{https://cloud.dwavesys.com/}); Microsoft (\href{https://azure. microsoft.com/en-us/services/quantum/}{https://azure. microsoft.com/en-us/services/quantum/}), Amazon (\href{https://aws.amazon.com/braket/}{https://aws.amazon.com/braket/}), and Google have also announced plans to enable cloud access to their processors. Google recently used a 53-qubit processor (named Sycamore) to report the first demonstration of quantum computational supremacy~\cite{arute_supremacy_2019}, defined as solving a problem using a quantum computer significantly faster than the best-known algorithm on a classical computer~\cite{Preskill2011}. Sycamore comprises 53 individually controllable transmon-type qubits and 86 couplers used to turn on/off nearest- neighbor 2-qubit interactions. The calibrated single-qubit and 2-qubit gate fidelities fell in the 99–99.9\% range when operating all qubits simultaneously. Google’s processor performed a task related to random number generation: sampling the output of a pseudorandom quantum circuit. The compatibility of this task with NISQ processors, along with the difficulty of simulating it classically, made it a promising candidate for a first quantum supremacy demonstration. The protocol is implemented by a sequence of operational cycles, each comprising randomly selected single-qubit gates and prescribed two-qubit gates applied to each qubit in the processor. At 53 qubits and 20 cycles, Google showed that their processor could perform the sampling in 200 seconds, outpacing the world’s most powerful supercomputer by between three~\cite{Pednault2019} and nine~\cite{arute_supremacy_2019} orders of magnitude, depending on the algorithm and hardware assumptions used for the classical computer. This outperformance of the best classical computing resources available represents a watershed achievement for the field of superconducting quantum computing

% % ############# 
% % FINALIZE UPDATING THIS PARAGRAPH AND SECTION LINKS IN THE INTRODUCTION TO GET THE NEW NUMBERING RIGHT!!!!!!
% 109: IBM News Room. 2016. IBM makes quantum computing available on IBM cloud to accelerate innovation.
% News release, May 4. https://www-03.ibm.com/press/us/en/pressrelease/49661.wss
% 110: AruteF,AryaK,BabbushR,BaconD,BardinJC,etal.2019.Nature574:505–10

% 111: PreskillJ.2011.Quantumcomputingandtheentanglementfrontier.Paperpresentedatthe25thSolvayCon-
% ference on Physics, The Theory of the Quantum World, Brussels, Belgium, Oct. 19–22. arXiv:1203.5813
% 112: PednaultE,GunnelsJA,NanniciniG,HoreshL,WisnieffR.2019.arXiv:1910.09534

\subsection{Quantum simulations with superconducting circuits}\label{sec:qsimulation}
% -------------------------------------------------------------------------------------
\begin{figure}[t]
\includegraphics{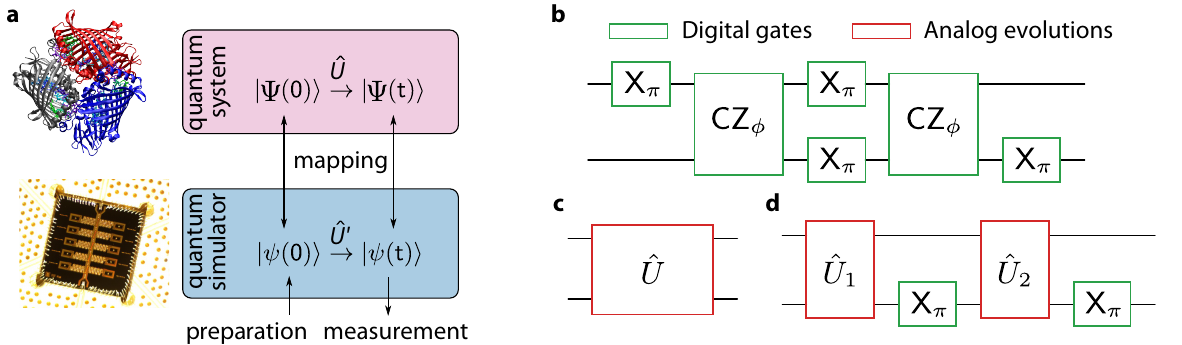}
\caption{Schematic representation of a quantum simulation system (a) A quantum system of interest (red, artistically depicted as a protein complex), is mapped onto an artificial quantum simulator (blue, here a superconducting circuit). By preparing and reading out quantum states $|\psi\rangle$ in the precisely controlled simulator -- therefore accessing its time evolution $\hat U '$ -- the time evolution $\hat U$ of the underlying quantum system of interest (described by states $\hat\Psi$) can be inferred. Figure adapted from Ref.~\cite{Georgescu2014}. (b) Pulse sequence used in a digital quantum simulation, comprised of one- and two-qubit gates used to construct the interaction Hamiltonian of a Fermi-Hubbard simulator~\cite{Barends2015}. (c) Analog quantum simulation: One-to-one mapping of the time evolution in the simulator and in the underlying model, see e.g. Ref.~\cite{Braumueller2017,Potocnik2018}. (d) Combination of analog unitary blocks and digital gates in the hybrid approach, as used in Ref.~\cite{Langford2017}.}
\label{fig:qsimulation}
\end{figure}

One of the most anticipated applications of quantum computation in the NISQ era is quantum simulation~\cite{Paraoanu2014}. A quantum simulator is a well-controllable device that mimics the dynamics or properties of a complex quantum system that is typically less controllable or accessible~\cite{Buluta2009}. The key idea is to study relevant quantum models by emulating or simulating them with hardware that itself obeys the laws of quantum mechanics~\cite{Feynman1982}, in order to avoid the exponential scaling of classical computational resources~\cite{Lloyd1996}.

Quantum simulators are problem-specific and do not meet the requirements of a universal quantum computer in general~\cite{Georgescu2014}. This simplification is reflected in the hardware requirements and may allow for a computational speed-up with few~\cite{Aspuru-Guzik2005}, even noisy quantum elements~\cite{Garcia-Ripoll2008}. Therefore, quantum simulations are likely to address meaningful problems with a quantum advantage well before universal quantum computation will be a reality \cite{Georgescu2014,Paraoanu2014}.

Certain qubit modalities are advantageous over others, as the qubits themselves may share intrinsic coupling mechanisms or commutation relations with the system to be simulated. The advantages of superconducting circuits for quantum simulation experiments are their high degree of control in manipulation, preparation, and efficient readout, together with the possibility to tailor circuit properties and implement tunable qubit frequencies and coupling strengths~\cite{devoret_superconducting_2013,Oliver2013}. The absence of intrinsic conservation laws when encoding abstract circuit excitations also makes superconducting circuits appealing for the study of non-equilibrium phenomena~\cite{Houck2012}.

Figure~\ref{fig:qsimulation}a schematically depicts the basic idea of a quantum simulation. The key requirement is an exact (or to a degree desired) mapping between the time evolution $\hat U$ of the quantum system of interest and the time evolution $\hat U '$ of the quantum simulator.
Two flavors of quantum simulations have been proposed and successfully demonstrated, coined digital and analog quantum simulation~\cite{Georgescu2014}. In the absence of quantum error correction (see Sec.~\ref{sec:QEC}), the achievement of a (problem specific) quantum advantage seems to be within closer reach for the analog or a hybrid analog-digital approach.

\subsubsection{Digital quantum simulation}
% -------------------------------------------------------------------------------------
Digital quantum simulation is a gate-based approach where a complex evolution is deconstructed into a set of single and two-qubit gates that can be implemented on the simulator hardware.
\begin{marginnote}[]
\entry{Lie-Trotter-Suzuki}{A commonly used method of decomposing the evolution under a local Hamiltonian into a sum of universal quantum gates (time steps) that can be efficiently implemented on quantum hardware.}
\end{marginnote}
It is closely related to universal quantum computation and is compatible with error correcting schemes (see Sec.~\ref{sec:QEC}). The approach relies on the fact that unitary operations that describe the time evolution of local Hamiltonians (that appear in most models of physical relevance) can be decomposed into a sum of local universal quantum gates~\cite{Lloyd1996}. The error introduced by the commonly employed Lie-Trotter-Suzuki decomposition~\cite{Suzuki1990} arises from non-vanishing commutators between the decomposed Hamiltonians and scales with the size of individual Trotter steps~\cite{Babbush2015}.
The digital simulation scheme was applied to study up to four fermionic modes with a superconducting quantum circuit~\cite{Barends2015}. Fermion operators with their correct anti-commutation relation were expressed in terms of Pauli operators using the Jordan-Wigner transformation, establishing an efficient mapping of the Fermi-Hubbard model to a spin Hamiltonian~\cite{Heras2015}.
\begin{marginnote}[]
\entry{Jordan-Wigner transformation}{One example of mapping fermionic operators to Pauli spin operators, enabling to simulate fermionic models with a set of (spin-like) qubits.}
\end{marginnote}
The gate sequence used to effectively construct the relevant interactions is depicted in Fig.~\ref{fig:qsimulation}(b). Spin models were investigated with a digital quantum simulation of an adiabatic algorithm on a nine-qubit chip~\cite{Barends2016} and a two-qubit chip~\cite{Salathe2015}, similarly using a Trotter decomposition in order to construct all interactions necessary to recover the model dynamics~\cite{Heras2014}. While these experiments highlight the versatility and universality of the digital approach, the total number of Trotter steps did not exceed ${\sim} 5$ due to the gate errors present in these systems. Several theory proposals address the efficient encoding of gate sequences in digital quantum simulators~\cite{Babbush2015,Garcia-Alvarez2016}, which, notwithstanding the hardware requirements, is one of the major challenges of this approach.
\begin{marginnote}[]
\entry{Hubbard models}{Highly general condensed matter model that describes a lattice of coupled fermions or bosons with an on-site interaction.}
\end{marginnote}

\subsubsection{Analog quantum simulation}
% -------------------------------------------------------------------------------------
In analog quantum simulations, the simulator directly mimics the time evolution of the quantum system of interest instead of constructing it, see Fig.~\ref{fig:qsimulation}(c). This requires a close mapping between system and simulator Hamiltonians in order to emulate the continuous time evolution, which in turn ensures a good scaling of hardware resources with problem complexity.
Circuit-based analog quantum simulators have been proposed recently for studying Andersen and Kondo lattices~\cite{Garcia-Ripoll2008}, Ising models~\cite{Viehmann2013} and phase transitions therein~\cite{Tian2010,Zhang2014}, fermionic models ~\cite{Reiner2016}, investigating Holstein polarons~\cite{Mei2013}, and for exploring relativistic quantum mechanics~\cite{Pedernales2013,Ballester2012}.
An array of coupled superconducting qubits naturally emulates the repulsive Bose-Hubbard model. Recently, 1D Bose Hubard chains were experimentally implemented to study quantum random walks~\cite{Yan2019} of one and two particles (excitations), and the stabilization of a Mott insulator phase~\cite{ma_dissipatively_2019}.
The study of quantum many-body effects is another application adopted by analog quantum simulation. By generating a synthetic magnetic field, a quantum phase exhibiting a chiral ground-state current was observed with a mutually coupled three-qubit unit cell~\cite{Roushan2016}, and many-body localization signatures were experimentally demonstrated by using a spectroscopy technique that maps out the eigenenergies of a Hamiltonian of interest~\cite{Roushan2017}.
Excitation transport in photosynthesis was recently studied by spectroscopic means on a three-qubit superconducting device~\cite{Potocnik2018}, inspired by an earlier proposal~\cite{Mostame2012}. Here, all temperatures and frequencies are scaled by ${\sim}10^5$ in order to establish a correct mapping to biological mechanisms.
Open quantum systems described by the spin boson model~\cite{Leggett1987} have recently attracted interest in the context of circuit simulations~\cite{Magazzu2018,Leppaekangas2018} for its straightforward mapping to superconducting qubits and resonators while being ubiquitous in nature and hard to access classically.
It reveals its complex quantum dynamics especially in the \textit{ultra-strong coupling} (USC) regime, which was simulated spectroscopically~\cite{Forn-Diaz2016} by implementing an increased physical coupling strength between a flux qubit and a transmission line. Several experiments investigated the non-classical groundstate properties of the quantum Rabi model at USC~\cite{Forn-Diaz2010,Yoshihara2016}, and the model dynamics were emulated by creating an effective quantum Rabi model at USC in a rotating frame~\cite{Braumueller2017,Ballester2012}. In contrast to atom- and ion-based qubit implementations, superconducting qubits strongly interact with electromagnetic fields, facilitating such schemes. While the analog approach equally suffers from the finite coherence of the simulator hardware, a noisy environment may be considered as part of the simulation, accounting for the natural noise channels in the physical system being simulated. It remains an open question how to correctly benchmark the performance of such a lossy analog quantum simulator.
\begin{marginnote}[]
\entry{Quantum Rabi model}{The quantum Rabi model describes a two-level atom or qubit coupled to a quantized harmonic mode via a transversal
interaction of arbitrary strength. At weak coupling, it can be reduced to the celebrated Jaynes-Cummings model.}
\entry{USC}{In the USC regime of the quantum Rabi model, the coupling strength is comparable to the resonator and qubit energies, leading to a breakdown of the rotating-wave approximation and consequently to dynamics that are hard to track classically.}
\end{marginnote}

\subsubsection{Digital-analog approach}
% -------------------------------------------------------------------------------------
A recent development is the strategy of analog-digital quantum simulations~\cite{Lamata2018}. By merging analog unitary blocks and digital gates, the overhead in gate construction is decreased and the advantageous scaling properties of the analog simulator are preserved, while the digital steps enhance the versatility of the simulator. This approach was used to simulate the quantum Rabi model at USC by constructing necessary Hamiltonian terms with digital gates while the simulation relied on an evolution in (different) analog blocks~\cite{Langford2017,Mezzacapo2014}. See Fig.~\ref{fig:qsimulation}(d) for the simplified basic Trotter step used in the experiment. The digital-analog approach was likewise used in a proposal of a fermion-fermion scattering experiment on a three-qubit superconducting circuit that comprises an open transmission line~\cite{Garcia-Alvarez2015}.

\subsection{Small-scale quantum algorithms}\label{sec:digitalquantumalgorithms}
% -------------------------------------------------------------------------------------
Digital quantum algorithms in the NISQ setting represent an interesting alternative to algorithms that rely on the full power of universal, error-corrected quantum computers. Early demonstrations of quantum algorithms in superconducting circuits focused on small, non-QEC versions of well-known quantum algorithms such as Deutsch-Jozsa \cite{dicarlo_demonstration_2009} and Shor \cite{Lucero2012}, as well as demonstrations of surface code primitives (discussed in Sec.~\ref{sec:QEC}).

NISQ algorithms may be thought of as `hardware-informed' quantum algorithms, i.e. the algorithms can be developed for a specific qubit connectivity to avoid certain low-fidelity qubits in the processor or hard-to-implement quantum gates. NISQ algorithms do not rely on the full support of quantum error correction, but instead optimize the algorithm fidelity based on an expectation of a lossy quantum system.  In particular, the most promising NISQ algorithms take hybrid classical-quantum approaches: they rely on classical computers and algorithms to implement the bulk of the necessary calculations, and tap the quantum processor only for the portions of the algorithm that cannot be performed efficiently on a classical processor. By delegating work to a classical processor, these algorithms reduce the circuit depth\begin{marginnote}[]
\entry{Circuit depth}{The number of time-steps required to run a given quantum algorithm. For instance, the depth of the quantum circuit in Fig.~\ref{fig:qsimulation}(b) is $5$.}
\end{marginnote} and therefore minimize the impact of circuit decoherence on the accuracy of the algorithm.

\subsubsection{Quantum chemistry}
% -------------------------------------------------------------------------------------
Quantum chemistry is potentially one of the `killer apps' for a quantum computer. This is due in large part to the development and demonstration of the variational quantum eigensolver (VQE), a hybrid classical-quantum algorithm that places relatively modest requirements on the quantum system  (see e.g. Ref.~\cite{moll_quantum_2018} and references therein). In the VQE, the Hamiltonian of a multi-atom system is mapped onto an array of qubits using an efficient classical algorithm, such that the Hamiltonian can be written as
\begin{equation}
\hat{\mathcal{H}} = \sum_{i\alpha} h^i_\alpha\hat\sigma ^i_\alpha + \sum_{ij\alpha\beta} h^{ij}_{\alpha\beta} \hat\sigma ^i_\alpha \hat\sigma ^j_\beta + ...,
\end{equation}
where $\{i,j, ...\}$ index over qubits, $\{\alpha, \beta, ...\}$ index over Pauli-matrix elements, and $\{h_{\alpha,\beta}^{ij}\}$ are classically computed weights \cite{Peruzzo2014}. To find the energy levels of the molecule, one initializes a test state on the quantum system, measures the relevant expectation values $\langle \hat\sigma ^i_\alpha \hat\sigma ^j_\beta\rangle$, and reconstructs the total energy $\langle \hat{\mathcal{H}} \rangle$ of the state. By using a classical minimization algorithm with $\langle \hat{\mathcal{H}} \rangle$ as the objective function, one can find an upper limit to the ground state energy. Once the ground state is known, the higher energy levels can be estimated using quantum subspace expansion (QSS) \cite{Mcclean2016} or an equation-of-motion (EOM) approach \cite{Rowe1968}.

The first VQE demonstration in a superconducting qubit system was performed by O'Malley et al., who demonstrated its effectiveness in using two qubits to map the ground state of the H$_2$ molecule as a function of inter-atomic spacing \cite{OMalley2016}.  Kandala et al. used similar methods with up to six qubits to map the ground states of larger molecules, including LiH and BeH$_2$ \cite{Kandala2017}. Colless et al. used the QSS to map the excited states of H$_2$ using a noise-resilient variant of the VQE \cite{Colless2018}; Ganzhorn et al. also calculated higher energy levels, using an efficient gate set to generate the ground state and the EOM method to extract excited state energies~\cite{Ganzhorn2018}.

\subsubsection{Data processing on quantum computers}
% -------------------------------------------------------------------------------------
There have been a number of important algorithmic developments and demonstrations related to data processing on quantum computers and quantum machine learning (QML). One canonical QML algorithm is the Harrow, Hassidim, and Lloyd (HHL) algorithm for sampling solutions to systems of linear equations \cite{Harrow2009}. This algorithm can in certain settings provide exponential speedup over its classical counterparts; a four-qubit implementation was demonstrated by Zheng et al. \cite{Zheng2017}. However, the HHL algorithm is not NISQ-optimized and makes rather stringent demands on the system's ability to store and manipulate coherent quantum information.

One of the most promising QML algorithms for the NISQ era is the Quantum Approximate Optimization Algorithm (QAOA) \cite{Farhi2014}. The QAOA provides an approximate solution to an NP-hard multivariate minimization problem in polynomial time, with a guaranteed accuracy set by the algorithm.  Like the VQE, the QAOA allows for the bulk of calculations to be performed in the classical processor, with the quantum device only required to produce a certain quantum state and perform a set of quantum measurements. The QAOA was first demonstrated by Otterbach et al. using a 19-qubit processor \cite{Otterbach2017}.

A third set of machine-learning NISQ algorithms relate to data classification.  These algorithms, which represent quantum equivalents of neural networks \cite{Farhi2018,Mitarai2018}, take advantage of variational techniques to enable both supervised and unsupervised data classification mechanisms. In particular, Havlicek et al. demonstrated two supervised learning algorithms using two qubits on a five-qubit processor and laid out the case for the potential existence of feature maps for which a provable quantum advantage could be demonstrated \cite{Havlicek2018}. On a similar five-qubit processor, Rist\'e et al. implemented the so-called `learning parity with noise' problem, which exhibits a quantum advantage \cite{Riste2017}.

However, the assumption of access to some form of quantum random access memory (QRAM) \cite{Giovannetti_2008} in several of the proposed schemes for analyzing classical data on quantum computers poses an open question for the feasibility of these protocols (see e.g. Ref.~\cite{ciliberto_carlo_quantum_2018} and references therein for details).\begin{marginnote}[]
\entry{QRAM}{A quantum form of random access memory which can store either quantum or classical data, but has the ability that stored data can be addressed using a superposition input.}
\entry{RAQM}{A random access quantum memory, where classical bits give the address of a quantum state to be retreived.}
\end{marginnote} Using a single parametrically driven transmon qubit, a form of random access quantum memory (RAQM) was demonstrated by Naik et al. \cite{naik_random_2017}. 

\subsection{Quantum annealing}\label{sec:quantumannealing}
% -------------------------------------------------------------------------------------
A formally equivalent approach to universal quantum computation is adiabatic quantum computation~\cite{Aharonov2008}, where the solution to computational problems is encoded into the ground state of a time-dependent Hamiltonian~\cite{Vinci2017}. Solving the problem translates into an adiabatic quantum evolution towards the global minimum of a potential energy landscape that represents the problem Hamiltonian. In classical annealing -- used as a general heuristic for solving optimization problems -- this is achieved by using simulated thermal fluctuations that allow the system to escape local minima, in combination with an appropriate annealing schedule that ensures the adiabaticity condition~\cite{Kadowaki1998}. In quantum annealing, transitions between states are caused by quantum fluctuations rather than thermal fluctuations, leading to a more efficient convergence to the groundstate for certain problems~\cite{Kadowaki1998,Boixo2016,hauke_perspectives_2019}. Quantum annealers strive to implement ideal adiabatic quantum computation for a restricted set of Hamiltonians, but suffer from experimental compromises~\cite{Kadowaki1998}, at the expense of universality or adiabaticity~\cite{Vinci2017}.

The most notable experimental implementation of quantum annealing to date is a device with ${\sim} 2000$ superconducting flux qubits manufactured by D-Wave (D-Wave Systems, Burnaby, Canada). Frequency tunable qubits are arranged in inter-coupled unit cells comprising eight qubits, where each qubit in a unit cell is longitudinally coupled to four other qubits in the final Hamiltonian, defining a so-called "Chimera" graph~\cite{Harris2018}. The D-Wave devices can model the transverse-field Ising Hamiltonian,
\begin{equation}
\hat H=\Lambda(t)\left[\sum_i h_i \hat\sigma _i ^z +\sum_{i<j}J_{ij}\hat\sigma _i ^z \hat\sigma _j ^z\right] +\Gamma(t)\sum_i \Delta_i\hat\sigma _i ^x,
\end{equation}
where $2h_i$ are the asymmetry energies, $J_{ij}$ are the coupling matrix elements, and $\Delta_i$ are the tunneling energies. At the beginning of the quantum annealing process, $\Gamma(0)=1$ and $\Lambda(0)=0$ in order to create a known groundstate, being an equal superposition in the computational basis. During the annealing protocol, $\Gamma$ is adiabatically ramped to zero while $\Lambda$ is increased to unity in order to adiabatically evolve to the final Ising Hamiltonian.

In a recent experiment~\cite{Harris2018}, a three-dimensional lattice of $512$ Ising spins was simulated on the D-Wave device in order to map out the magnetic phase diagram of a spin glass. In a similar experiment, the D-Wave group studied the Kosterlitz-Thouless phase transition in a frustrated Ising model~\cite{King2018}. Both papers demonstrate that a variety of relevant lattices are accessible to the D-Wave approach by using non-trivial encodings, enabling research of condensed matter phenomena that are hard to address classically at a large scale. Recently, a framework was developed that maps the prime factorization problem on the D-Wave Ising model~\cite{Jiang2018}, demonstrating a reduced cost of $\mathcal{O}(\log^2(N))$ qubits (where $N$ is the integer number to be factorized). This has led to the experimental factorization of a seven-digit number with 89 qubits on the D-Wave machine~\cite{peng_factoring_2019}.
A definitive demonstration of a quantum enhancement for a general class of problems has been elusive for the D-Wave machines and quantum annealing in general~\cite{Ronnow2014}. In this context, it is known that the current D-Wave architecture is only able to implement stoquastic Hamiltonians, which can oftentimes be simulated efficiently with classical algorithms~\cite{Vinci2017,Bravyi2008}. \begin{marginnote}[]
\entry{Stoquasticity}{A Hamiltonian is stoquastic when its groundstate can be expressed as a classical probability distribution, allowing for more efficient classical sampling due to the absence of the "sign problem"~\cite{Bravyi2008}. Non-stoquastic Hamiltonians are believed to be inefficient to simulate classically.}
\end{marginnote}Recent studies, however, indicate that the D-Wave machine achieves significant runtime advantages for a certain class of problems~\cite{Denchev2016,Mandra2016}. It is an open question whether this is due to a quantum speedup or corresponds to a more efficient classical computation.

D-Wave recently demonstrated a quantum annealing experiment of a non-stoquastic Hamiltonian on a two flux qubit chip with fixed capacitive transversal coupling \cite{ozfidan_demonstration_2019}. Another experiment has been demonstrated that simulates non-stoquastic Hamiltonians on a nine-transmon qubit chip~\cite{Barends2016}, where non-stoquasticity was created by incorporating digital gates that construct the necessary distinct couplings.

% SECTION 4: QUANTUM ERROR CORRECTION AND FAULT-TOLERANCE
% -------------------------------------------------------------------------------------
\section{QUANTUM ERROR CORRECTION WITH SUPERCONDUCTING QUBITS}\label{sec:QEC}
Despite the tremendous progress on coherence, gate operations, and readout fidelity achieved with superconducting qubits, quantum error correction (QEC) will still be needed to realize truly large-scale quantum computers. 
Most QEC schemes utilize some form of redundancy (typically, multiple qubits) to encode so-called logical qubits. A prescription for performing the encoding and for correcting errors in the encoding is referred to as an error correcting code. The threshold theorem~\cite{gottesman_stabilizer_1997,aharonov_fault-tolerant_2008} then guarantees that for a QEC code, if the operational error-rate on the physical qubits is below a certain value, and the code is implemented in a fault-tolerant manner (see Sec.~\ref{subsec:FT}), then errors can be suppressed to arbitrary precision (see e.g. Ref.~\cite{nielsen_quantum_2011} for a general introduction to QEC). The two-dimensional surface code is perhaps the most promising, experimentally feasible QEC code in the near term, due to its particularly lenient error rate to satisfy the threshold theorem (error rate $\lesssim 1\%$), and because it only requires weight-four parity measurements using nearest-neighbour coupling to four qubits (see e.g. Refs.~\cite{fowler_surface_2012,tomita_low-distance_2014} and references therein for details). \begin{marginnote}[]
\entry{Logical qubit}{A redundantly encoded qubit in which quantum errors in the constituent components can be identified and corrected without corrupting the encoded qubit. A logical qubit beyond the `break-even point' has improved coherence (and potentially gate operation properties) over its uncorrected components.}
\end{marginnote}As a consequence, much of the experimental progress towards QEC has been focused on realizing multi-qubit parity measurements as well as primitives towards the surface code.
\begin{textbox}[t]\section{Parity measurements - a workhorse in quantum error detection and correction}
Many quantum error correction schemes rely on parity measurements. In the left circuit below, the ancilla qubit $|A\rangle$ is used to infer the bit parity (via information about $\langle Z_1Z_2\rangle$) of the two data qubits in state $|\Psi\rangle$, and in the right circuit qubit $|A\rangle$ infers the phase parity (via $\langle X_1X_2\rangle$),
$$
\Qcircuit @C=0.75em @R=1em{
 \lstick{\raisebox{-0.6cm}{$|\Psi\rangle$}\,} &	\lstick{\raisebox{-0.75cm}{$\left\{ \vphantom{\rule{0.1cm}{0.3cm}}\right.$}} & 	\ctrl{2}	& \qw		      & \qw & \qw & \\
 											&  	& 	 \qw		& \ctrl{1}		  & \qw & \qw & \\
 												& \lstick{|A\rangle} 	& 	 \targ 		& \targ &  \meter & \cw &     & \quad \langle Z_1Z_2 \rangle_\Psi\\
}
\quad\quad\quad\quad\quad
\raisebox{-0.55cm}{\text{and}}
\quad\quad\quad\quad\quad
\Qcircuit @C=0.75em @R=0.5em{
 \lstick{\raisebox{-0.6cm}{$|\Psi\rangle$}\,} &	\lstick{\raisebox{-0.75cm}{$\left\{ \vphantom{\rule{0.1cm}{0.3cm}}\right.$}} & \qw 		& \targ		& \qw	 	& \qw &\qw & \qw & \\
 												&  																			& 	\qw 	& \qw		& \targ		& \qw &\qw & \qw & \\
 												& \lstick{|A\rangle} 														& 	\gate{H}& \ctrl{-2}	& \ctrl{-1} & \gate{H} & \meter & \cw &     & \quad \langle X_1X_2 \rangle_\Psi\\
}
$$
In the absence of errors on the ancilla qubit, the eigenvalue of $|A\rangle$ will contain information reflecting whether the two-qubit state $|\Psi\rangle$ is an eigenstate of $Z_1Z_2$ (or $X_1X_2$) with eigenvalue $+1$ or $-1$ without collapsing the state of the individual qubits in $|\Psi\rangle$. Since the operators $Z_1Z_2$ and $X_1X_2$ (and even multiples of more $Z$ and $X$ operators) commute, combinations of parity measurements across a larger grid of qubits can therefore be used to infer if and where a bit- or phase-flip error ocurred, without collapsing the underlying quantum data. The collection of ancilla qubit measurements is typically referred to as the syndrome of the error, and inferring the underlying error is known as decoding.

\end{textbox}
\subsection{Progress in error detection and correction using parity measurements} \label{subsec:surfacecode}
\begin{figure}[t]
\includegraphics[width=6.3in]{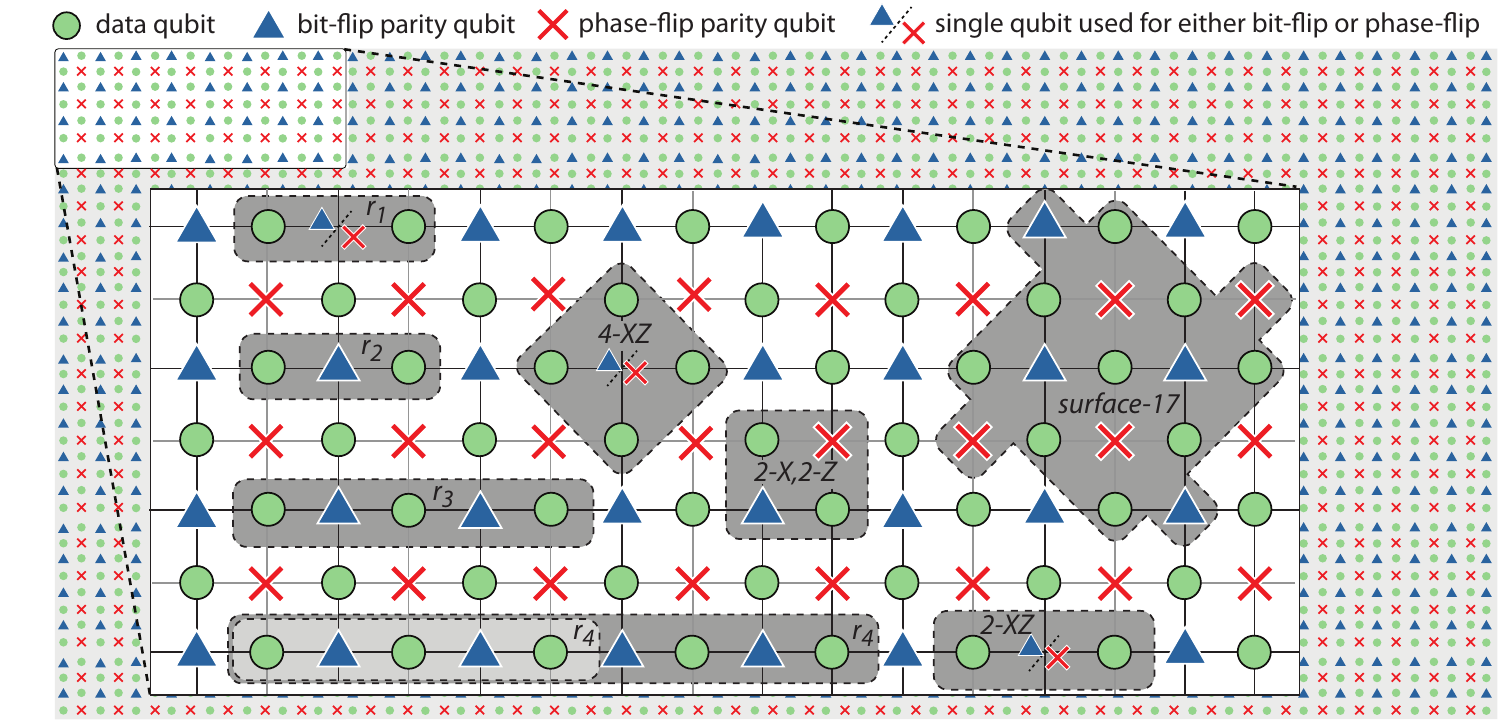}
\caption{A section of the qubit layout of the surface code, with $40\times 20$ data qubits (shown as circles), and associated bit-flip and phase-flip parity qubits are shown as triangles and crosses, respectively. Inset shows a subsection, in which shaded areas indicate parity experiments that have been reported, except `surface-17' which is currently being pursued in multiple laboratories (see text for details). Experiment $r_1$ by Reed et al.~\cite{reed_realization_2012}, $r_2$ by Chow et al.~\cite{chow_implementing_2014}, $r_3$ by Rist\'e et al~\cite{riste_detecting_2015}, $r_4$ by Kelly et al.~\cite{kelly_state_2015},  2-$X,$2-$Z$ by C\'orcoles et al.~\cite{corcoles_demonstration_2015}, 2-$XZ$ by Andersen et al.~\cite{andersen_entanglement_2019} and Bultink et al.~\cite{bultink_protecting_2019}, 4-$XZ$ by Takita et al.~\cite{takita_demonstration_2016}.}
\label{fig:surface-code-experiments}
\end{figure}

% -------------------------------------------------------------------------------------
Most experiments using superconducting qubits for quantum error detection and correction rely on parity measurements of two or more `data qubits', by coupling them to `syndrome' qubits. This basic construction has been used to demonstrate multiple aspects of error detection and error correction, which we review below. Figure~\ref{fig:surface-code-experiments} shows a section of the surface code, where circles correspond to the data qubits, and cross and triangle shapes are the ancilla syndrome qubits, used to infer the overall bit- and phase--parity of the neighboring data qubits. For a brief description of parity measurements, see the infobox on \textit{parity measurements} below. A general introduction to parity measurements (syndrome measurements) in the context of the surface code can be found in Ref.~\cite{fowler_surface_2012}.

Within the last 10 years, multiple experiments using superconducting qubits have realized parity measurements relying on a qubit layout corresponding to various sub-sections of the surface code. In particular, a single one-dimensional row of the surface code fabric, the so-called repetition code, which corrects either bit-flip or phase-flip errors, has been implemented in a multitude of ways. Reed et al.~\cite{reed_realization_2012} first demonstrated a three-qubit repetition code (corresponding to the section denoted $r_1$ and highlighted in the inset of Fig.~\ref{fig:surface-code-experiments}) using the \CZ{}-gate, together with the three-qubit controlled-controlled-phase gate (the \textsf{TOFFOLI}-gate) to correct a single error. Using the `cross-resonance' two-qubit gate, Chow et al.~\cite{chow_implementing_2014} demonstrated entanglement across three qubits (section $r_2$ in Fig.~\ref{fig:surface-code-experiments}). By using the middle qubit as a parity meter of the outer qubits, the authors were able to generate either the odd or even Bell states between the two non-nearest neighbor qubits, conditioned on the parity readout. Similar ideas were demonstrated simultaneously by Saira et al.~\cite{saira_entanglement_2014}.
Using an optimized form of the \CZ-gate~\cite{kelly_optimal_2014}, Barends et al.~\cite{barends_superconducting_2014} demonstrated a two-qubit gate fidelity of $\mathcal F \sim 0.994$ (see Tbl.~\ref{tab:2qubitgates}), using the `xmon' variant of the transmon qubit~\cite{Barends2013}. This was the first demonstration of a two-qubit gate in superconducting qubits whose fidelity (as measured via interleaved Clifford randomized benchmarking) surpassed the error threshold for the surface code. Kelly et al.~\cite{kelly_state_2015} then used this implementation of the \CZ-gate to demonstrate both a five- and nine-qubit repetition code (sections denoted $r_4$ in Fig.~\ref{fig:surface-code-experiments}). The authors performed multiple rounds of error detection, and then used a minimum-weight perfect-matching algorithm in post-processing to determine and correct the most likely physical errors. This process resulted in an improved encoded state fidelity, when going from a five-qubit repetition code to a nine-qubit code, showing for the first time the efficacy of the repetition code. As of writing, the experiments performed by Kelly et al. represent the largest (by qubit-count) repetition code to have been experimentally demonstrated. Work by Rist\'e et al.~\cite{riste_detecting_2015} studied performance of the bit-flip repetition code by artificially injecting coherent and incoherent noise after encoding a 3-qubit logical state (corresponding to the section denoted $r_3$ in Fig.~\ref{fig:surface-code-experiments}).

The repetition code by itself cannot simultaneously detect both bit- and phase-flip errors, so it cannot serve as a full quantum error correcting code. However, first work towards demonstrating full quantum error detection (as a precursor to full quantum error correction) of bit and phase flips was demonstrated by C\'orcoles et al.~\cite{corcoles_demonstration_2015}. By using a $2\times 2$ `half-plaquette' of the surface code (corresponding to  section 2-$X$,2-$Z$ in Fig.~\ref{fig:surface-code-experiments}), the authors demonstrated quantum error detection in a two-qubit Bell state by reading both the $Z_1Z_2$ and $X_1X_2$ parities within one round of error correction. 

Recently, stabilization of Bell states has been studied using a single ancilla to perform both $Z_1Z_2$ and $X_1X_2$ parity checks, corresponding to the region denoted 2-$ZX$ in Fig.~\ref{fig:surface-code-experiments}. The work by Andersen et al.~\cite{andersen_entanglement_2019} demonstrated real-time stabilization (using fast feedback) to maintain a Bell state fidelity of ${\sim} 0.74$ in up to 12 cycles of feedback. The experiment by Bultink et al.~\cite{bultink_protecting_2019} used the so-called "Pauli frame updating" technique to keep track of parity flips in up to $26$ rounds (corresponding to roughly \SI{20}{\micro s} experiment time) with a resulting state fidelity ${\sim} 0.8$.

However, unlike the repetition code experiments, the surface code relies on weight-four parity measurements, i.e., measuring operators of the form $Z_1Z_2Z_3Z_4$ or $X_1X_2X_3X_4$. The first demonstration of weight-four parity measurements in superconducting qubits was performed by Takita et al.~\cite{takita_demonstration_2016}, corresponding to the section labelled $4-XZ$ in Fig.~\ref{fig:surface-code-experiments}. By utilizing an optimized gate implementation to cancel spurious cross-talk issues related to the two-qubit cross-resonance gate, the authors achieved a weight-four parity fidelity of 0.774 for $Z_1Z_2Z_3Z_4$ and 0.795 for $X_1X_2X_3X_4$.

Finally, the smallest logical qubit that uses the surface code encoding and can be error corrected is denoted `surface-17' and is shown in the inset of Fig.~\ref{fig:surface-code-experiments}. The `surface-17' logical qubit uses nine data qubits and eight parity measurement qubits to simultaneously correct both bit- and phase-flip errors. Such a device is under current investigation using the \CZ{}-gate to implement two-qubit operations \cite{versluis_scalable_2017}.

\subsection{Fault tolerance using superconducting qubits} \label{subsec:FT}
% -------------------------------------------------------------------------------------
The notion of fault tolerance (FT) is a key component for realizing a scalable quantum computer from faulty physical components. FT is an architectural property of a QEC quantum circuit, which (roughly stated) ensures that single physical errors in the underlying components do not propagate in a single time-step to corrupt logical data (see e.g. Ref. \cite{gottesman_introduction_2009} for details).

Demonstrating explicit FT in superconducting qubit systems has to date been focused on using a version of the $[\![4,2,2 ]\!]$--code (see Ref.~\cite{gottesman_quantum_2016} and references therein), which requires five physical qubits in total. This code encodes two logical qubits into four physical qubits and uses a fifth qubit for error-detection. The low physical qubit count for this code comes at the expense of not being an error-correcting code, but only an error-detection code. In the experiments of Ref.~\cite{takita_experimental_2017} one FT encoded qubit, and one non-FT encoded qubit were initialized (limited by the connectivity of the device). In the presence of noise, the FT-encoded circuits were shown to produce the intended state with greater probability than the non-FT circuit.\begin{marginnote}[]
\entry{$[\![n,k,d ]\!]$--codes}{A shorthand notation for a quantum error correcting code using $n$ physical qubits (not counting overhead for parity readout or FT circuits), which encodes $k$ logical qubits, with a distance $d=2t+1$ that can correct $t$ errors.}
\end{marginnote} Vuillot also studied FT state preparation circuits~\cite{vuillot_is_2018}. By using the highest-quality pair of physical qubits on the device to generate a set of specific states, Vuillot was able to show that by preparing the same states, but with FT encoding circuits, led to an average improvement in the state preparation fidelity.

To deconvolve the effects of state preparation and measurement (SPAM) errors from the improvements due to FT encodings, Harper \& Flammia recently performed Clifford randomized benchmarking using the $[\![4,2,2 ]\!]$--code \cite{harper_fault-tolerant_2019}. The infidelity of the logical gates decrease by nearly an order of magnitude when FT encodings are used, relative to non-FT physically equivalent gates.

\subsection{Bosonic codes with superconducting cavities} \label{subsec:catcodes}
% -------------------------------------------------------------------------------------
Bosonic codes -- those built from the internal states of a QHO -- represent a different approach to demonstrating error-resilient qubits. Rather than encoding a logical qubit in the shared state of many two-level systems, as in the surface code, a bosonic code constructs a logical qubit from the many energy levels of a single quantum object (in this context, typically the long-lived photonic states of a superconducting cavity). Bosonic codes have had a remarkably rapid development trajectory and are now leading the superconducting qubit field in early prototype demonstrations of fault-tolerant error-corrected quantum computing. A few broad categories of bosonic codespaces have been considered, namely:

\textbf{\textit{{Fock-state encodings:}}} Qubits are mapped onto the $|0/1\rangle$ Fock states and form a direct analogy to a traditional spin 1/2 qubit encoding. Fock states are the longest-lived QHO qubits, but do not implement a bosonic code. \begin{marginnote}
\entry{Example cat codes}{
The `2-cat code': $|{0/1}\rangle_L = |{\alpha}\rangle \pm |{-\alpha}\rangle$
The `4-cat code': $|{0/1}\rangle_L = |{\alpha}\rangle + |{-\alpha}\rangle \pm |{i\alpha}\rangle \pm |{-i\alpha}\rangle$
}
\end{marginnote}
\begin{marginnote}
\entry{Example binomial codes}{Encoding 1:\\
$|{0}\rangle_L = |{2}\rangle$ \\$|{1}\rangle_L = \frac{|{0}\rangle + |{4}\rangle}{\sqrt{2}}$
Encoding 2:\\
$|{0/1}\rangle_L = \frac{1}{\sqrt{2}} \left( \frac{|{0}\rangle + |{4}\rangle}{\sqrt{2}} \pm |{2}\rangle\right)$}
\end{marginnote}Instead, they are the benchmark against which to evaluate the QEC logical qubit lifetime.

\textbf{\textit{{Cat codes \cite{Cochrane1999,Mirrahimi2014}:}}} Qubits are formed using superpositions of coherent states. The superpositions are chosen such that the loss of a single photon maps the system onto a detectable and correctable error space while maintaining the encoded information. Cat codes require a relatively large $\overline{n}$ in order to approach orthogonality in the codespace. With sufficiently high $\overline{n}$, error correction can be performed at the end of the experimental cycle as long as error detection is performed quasi-continuously.

\textbf{\textit{{Binomial codes \cite{Michael2016}:}}} Qubits are defined via a finite number of Fock states with binomial coefficients. Binomial codes comprise exactly orthogonal logical states constructed such that photon loss maps the system onto a correctable error space. The codespace can be constructed to allow for multiple photon losses at the expense of higher $\overline{n}$. Binomial codes are in some sense a hybrid between Fock encodings, which are not correctable, and cat states, which require larger $\overline{n}$ and are not exactly orthogonal. QEC on a binomial code requires correction at the time of detection.

Recent demonstrations have confirmed the potential of bosonic codes in QEC and for fault-tolerant quantum computing. The initial demonstration of mapping an arbitrary qubit state onto a cat state was performed in 2013 by Vlastakis et al. \cite{Vlastakis2013}. Error detection was demonstrated by observing parity jumps \cite{Sun2014}, with fidelity limited by the relatively short lifetime of the transmon ancilla qubit used to read and control the cavity state. Universal control of cavity states in general \cite{Heeres2015} and of logical bosonic codes in particular \cite{Heeres2017} were demonstrated, enabling measurement-based QEC of cat codes \cite{ofek_extending_2016} and binomial codes \cite{hu_demonstration_2018}. The former demonstration surpassed the break-even point by showing an error-corrected lifetime $T_{1, \text{QEC}}$ greater than the Fock state lifetime $T_{1,\text{Fock}}$; the latter came very close to this threshold and additionally demonstrated high-fidelity (although not error-corrected) operations on the logical single qubits.

Beyond single logical qubit demonstrations, there have been several recent demonstrations of entanglement generation and two-qubit gates between the encoded bosonic qubit states. These demonstrations include a logical \textsf{CNOT} \cite{rosenblum_fault-tolerant_2018}, \textsf{CNOT} gate teleportation between two error-correctable qubits machined from the same aluminum block \cite{Chou2018}, logical state transfer and remote entanglement between qubits separated by a long delay cable \cite{Axline2018} and an exponential-SWAP operation \cite{Gao2018}. While these multi-qubit experiments do not yet include active error correction, they have shown the feasibility of operations on logical states, and represent continuing progress towards universal QEC using superconducting hardware.

\section{LOOKING AHEAD}\label{sec:outlook}
% -------------------------------------------------------------------------------------
In this review, we have discussed several of the most recent advances in the development of qubit architectures, gate operation, amplification and readout, and digital- and analog-algorithm implementations, as well as work toward quantum error correction. Despite the already tremendous progress outlined in this review, the field is still undergoing rapid development, and the theoretical, experimental, and conceptual boundaries are consistently being pushed. Below, we briefly discuss some future directions and near-term challenges for the field.

\subsection{Beyond the surface code for quantum error correction} While the surface code is promising due to its relatively lenient error threshold and modest requirements on connectivity, the overhead of physical-to-logical qubits is daunting, and the fault-tolerant gate-set is limited~\cite{fowler_surface_2012}. Other topological and concatenated codes typically have more demanding error thresholds and connectivity requirements (see e.g.~\cite{Campbell2017a} and references therein for a more detailed discussion), but allow for fault-tolerant implementation of a larger gate set. Whether other codes will be experimentally feasible in the near term hinges to some extent on whether improving overall gate fidelity or improving qubit count and connectivity is the more difficult endeavor.

For the bosonic codes, while the preliminary experiments have been promising, significant challenges remain. Fault-tolerant single-qubit error detection has been demonstrated \cite{rosenblum_fault-tolerant_2018}, but there are no current proposals for fault-tolerant multi-qubit protocols. In the absence of fault-tolerant gates, universal computing will require embedding the bosonic qubit in a larger error-correcting fabric (such as a surface code), potentially implying all of the associated scaling issues just discussed.

Entirely different schemes rely on native error-resilience, providing another path forward for high-fidelity quantum memories and operations. Examples of this approach include the $0-\pi$ qubit~\cite{brooks_protected_2013}, `error-transparent' encodings~\cite{kapit_error-transparent_2018}, and the metastable flux qubit~\cite{kerman_2010}.

\subsection{Alternative superconducting qubits} The transmon qubit modality has shown tremendous progress over the last decade, but it has certain limitations. The charge noise resilience of the transmon qubit comes at the expense of a small anharmonicity, making it more likely for excitations to leave the computational subspace. This shortcoming is addressed by the fluxonium qubit~\cite{Manucharyan2009}, which can be considered as combining the advantages of the Cooper-pair box and the flux qubit, while avoiding their respective drawbacks, at the cost of introducing high numbers of Josephson junctions per qubit. While recent experiments indicate excellent coherence properties of fluxonium qubits~\cite{Pop2014,Nguyen2018}, it has not yet been used in larger scale, more complex circuits since its operation is less straightforward and its coupling capabilities in a circuit QED environment remain to be demonstrated.

A different strategy, which still relies on the transmon qubit modality, replaces the local flux control used in the tunable transmon qubits with local voltage control, by using superconductor-semiconductor-superconductor Josephson junctions. In such systems, a local electrostatic gate is used to tune the carrier density in the semiconductor region, resulting in a modified $E_\text{J}$. Such devices were first demonstrated in InAs nanowires proximitized by epitaxially-grown aluminum \cite{Larsen2015_1,de_lange_realization_2015}, forming the transmon qubit element in a cQED setup. Subsequently, improved coherence times as well as compatibility with large external magnetic fields were demonstrated~\cite{Luthi2018}. However, the need to individually place nanowires makes the path to larger devices within this scheme potentially difficult. Alternative demonstrations of such hybrid superconducting qubit systems have therefore used two-dimensional electron gases~\cite{Casparis2018} amenable to top-down fabrication, as well as graphene flakes proximitized by evaporated aluminum~\cite{Wang2019}. The absence of local currents results in a decrease of the power that needs to be delivered onto the qubit chip, but at the cost of reintroducing some charge noise susceptibility through the gate.

\subsection{Next steps}  
Although there is ample daylight ahead for both NISQ-era demonstrations and large-scale, FT quantum computers based on superconducting qubits, there are also many nontrivial obstacles to overcome. On the path toward large quantum processors, the demonstration of multiple error- corrected FT logical qubits with gate fidelities and lifetimes exceeding any of the constituent degrees of freedom will be an important step. Finally, we outline a few of the challenges facing the community, as quantum processors are now moving from 10–20-qubit scale to the 50–100-qubit scale.
\newpage
\begin{issues}[NEAR-TERM CHALLENGES]
\begin{enumerate}
\item[$\circ$] \textbf{Control and high coherence in medium-scale devices}: For medium- and large-scale devices, the individual qubit coherences are not necessarily the same as those in a simpler few-qubit devices. Maintaining high coherence and high-fidelity control across a large chip is a key challenge.
\item[$\circ$] \textbf{Scalable calibration techniques}: Advanced software strategies are also needed to calibrate medium-to-large scale quantum processors due to the large number of non-trivial cross-calibration terms while finding simultaneous optimal operating parameters.
\item[$\circ$] \textbf{Verification and validation}: As the number of qubits increases, efficiently determining the fidelity of quantum operations across the entire chip using e.g. Clifford randomized benchmarking~\cite{magesan_scalable_2011} becomes infeasible and new techniques for validation and verification will be needed. Techniques such as `cross entropy benchmarking'~\cite{boixo_characterizing_2018} and `direct benchmarking'~\cite{proctor_direct_2019} have recently been proposed and implemented.
\item[$\circ$] \textbf{Improving qubit connectivity}: While impressive progress has been made in three-dimensional integration of superconducting
circuits (e.g. Ref.~\cite{rosenberg_3d_2017}), non-planar connectivity of high-fidelity qubits has yet to be demonstrated.
\item[$\circ$] \textbf{Improved gate fidelity}: Continued improvements to gate fidelities will be an important step towards bringing down the overhead of physical qubits needed to encode a single logical qubit as well as important for demonstrating the efficacy of NISQ algorithms.
\item[$\circ$] \textbf{Robust \& reproducible fabrication}: The fabrication of medium-to-large scale superconducting circuits will need to be consistent with continued improvements to qubit coherence and 3D integration techniques.
\end{enumerate}
\end{issues}
Using current techniques -- nothwithstanding the challenges outlined above -- it seems possible to scale to on the order of $\sim$1000 qubits. However, beyond this (rough) number, a new set of techniques will be needed. Examples include co-location inside the dilution refrigerator of control and readout electronics, as well as on-the-fly decoders for quantum error correction procedures.

% End matter:
% -------------------------------------------------------------------------------------

% Disclosure
\section*{DISCLOSURE STATEMENT}
The authors are not aware of any affiliations, memberships, funding, or financial holdings that
might be perceived as affecting the objectivity of this review.

% Acknowledgements
\section*{ACKNOWLEDGMENTS}
The authors gratefully acknowledge input and comments from Laura García-Álvarez, Gabriel Samach, Antti Veps\"al\"ainen and Serge Rosenblum on the manuscript as well as numerous fruitful discussions with Terry P. Orlando. MK gratefully acknowledges support from the Carlsberg Foundation during part of this work. PK acknowledges partial support by the Wallenberg Centre for Quantum Technology (WACQT) funded by Knut and Alice Wallenberg Foundation. This research was funded in part by the U.S. Army Research Office Grant Nos. W911NF-14-1-0682, W911NF-18-1-0116, W911NF-18-1-0411, and MURI W911NF-18-1-0218; the National Science Foundation Grant No. PHY-1720311; and the Office of the Director  of National Intelligence (ODNI), Intelligence Advanced Research Projects Activity (IARPA) and the Assistant Secretary of Defense for Research and Engineering via MIT Lincoln Laboratory under Air Force Contract No. FA8721-05-C-0002. Opinions, interpretations, conclusions, and recommendations are those of the authors and are not necessarily endorsed by the United States Government.

% Bibliography
%%%%%%%%%%%%%%%%%%%%
\bibliographystyle{ar-style4}

\end{document}